\documentclass[lettersize,journal]{IEEEtran}
\usepackage{amsmath,amsfonts,amssymb}
\usepackage{algorithm}
\usepackage{array}
\usepackage{textcomp}
\usepackage{stfloats}
\usepackage{url}
\usepackage{verbatim}
\usepackage{graphicx}
\usepackage{color}
\usepackage{multirow}
\usepackage{makecell}
\usepackage{booktabs}
\usepackage{subfigure}
\usepackage{csquotes} 
\usepackage{cite}
\usepackage{algpseudocode}

\begin{document}
\bstctlcite{IEEEexample:BSTcontrol}

\title{A Novel Spreading-Factor-Index-Aided LoRa Scheme: Design and Performance Analysis
}

\author{
    \IEEEauthorblockN{Hao Zeng, Huan Ma, \textit{Member, IEEE}, Yi Fang, \textit{Senior Member, IEEE},
    \\ Pingping Chen, \textit{Senior Member, IEEE}, Wenkun Wen, and Tierui Min}
    \thanks{This work was supported in part by the NSF of China under Grants 62322106 and 62571142. \textit{(Hao Zeng and Huan Ma contribute equally to this work.)}}
    \thanks{Hao Zeng and Yi Fang are with the School of Information Engineering, Guangdong University of Technology, Guangzhou 510006, China (e-mail: haozeng2025@163.com; fangyi@gdut.edu.cn).}
    \thanks{Huan Ma is with the Electronics and Information Engineering Department, Zhaoqing University, Zhaoqing 526061, China (e-mail: mh-zs@163.com).}
    \thanks{Pingping Chen is with the Department of Electronic Information, Fuzhou University, Fuzhou 350116, China (e-mail: ppchen.xm@gmail.com).}
    \thanks{Wenkun Wen and Tierui Min are with Techphant Technologies Guangzhou Ltd. Co., Guangzhou 510000, China (email: wenwenkun@techphant.net, mintierui@techphant.net).}
}

\maketitle

\begin{abstract}
LoRa is a widely recognized modulation technology in the field of low power wide area networks (LPWANs). However, the data rate of LoRa is too low to satisfy the requirements of Internet of Things applications. To address this issue, we propose a novel high-data-rate LoRa scheme based on the spreading factor index (SFI). In the proposed SFI-LoRa scheme, the starting frequency bin of a chirp signal is used to transmit information bits, while the combinations of spreading factors are exploited as a set of indices to convey additional information bits. Moreover, the theoretical symbol error rate, data rate, transmission throughput, complexity and energy efficiency of the proposed SFI-LoRa scheme are carefully analyzed. Simulation results not only verify the accuracy of our theoretical analysis, but also demonstrate that the proposed SFI-LoRa scheme can improve the transmission throughput of existing LoRa schemes without sacrificing the BER performance over additive white Gaussian noise, Rayleigh fading, and multipath flat-fading channels. Therefore, the proposed SFI-LoRa scheme is a potential solution for applications requiring a high data rate in the LPWAN domain.
\end{abstract}


\section{Introduction}
\IEEEPARstart{I}NTERNET of Things (IoT) technologies are progressively becoming an indispensable component of our lives. A forecast from International Data Corporation estimates that there will be 41.6 billion connected IoT devices and generating 79.4 zettabytes of data in 2025 \cite{ref26}. With the continuous development of 5G technology and the rapid growth in the number of devices connected to the internet, IoT has become the milestone in the evolution of digital strategy \cite{ref51,ref64}. As a solution drawn a significant amount of research attention, the deployment of low-power wide area networks (LPWANs) has been extended across numerous fields \cite{ref2,ref65,ref66, ref90}.

The three most popular LPWAN technologies in the IoT are NB-IoT \cite{ref29}, Sigfox \cite{ref30}, and LoRa \cite{ref31}. In the above LPWAN technologies, LoRa has been widely adopted because it allows to build and maintain an autonomous network without third-party infrastructure \cite{ref91}. Compared to NB-IoT, which relies on cellular infrastructure, LoRa provides wider network coverage without the expensive cellular subscriptions. Although Sigfox operates on a narrowband approach, it is tied to a proprietary network, whereas LoRa is more scalable and accessible with an open standard \cite{ref52}. The flexibility and cost-effectiveness make LoRa a more attractive solution for many IoT applications. From the technical perspective, LoRa-based LPWAN can be categorized into two layers. The physical layer and medium access control layer of LoRa  use chirp spread spectrum (CSS) modulation and ALOHA-based protocol \cite{ref34}, respectively. In the LoRa modulation, a multidimensional space for LoRa signals is established through the utilization of cyclic shifts of a chirp signal, characterized by a linear increase in frequency. It is crucial to note that different LoRa signals exhibit mutual orthogonality \cite{ref4}, \cite{ref32}. The frequency of these LoRa signals has been observed to increase linearly at different chirp rates, with the increasing rate being dependent upon the spreading factor (SF) ($f=7,8,...,12$) \cite{ref14,ref33}. Increasing the SF is capable of expanding the LoRa coverage area. However, it is also recognized that this can result in a reduction in the data rate.

%
%
%
%
%
%

\begin{table}[t]
\centering
\caption{\textsc{Comparison of the Five IoT Solutions: NB-IOT, Sigfox , Redcap, eRedcap, and LoRa.}}

\label{tab0}
\begin{tabular}{|m{1.5cm}<{\centering}|m{1.2cm}<{\centering}|m{1.2cm}<{\centering}|m{1.2cm}<{\centering}|m{1.2cm}<{\centering}|}

\hline
Criterion &  Frequency band &   Data rate  & Power consumption  & Coverage \\ \hline

NB-IoT  &  $800~\rm{MHz}$, $900~\rm{MHz}$, $1800~\rm{MHz}$ & $26~\rm{kbps}-66~\rm{kbps}$  & Low  & Long\\ \hline

Sigfox  &   $868~\rm{MHz}$, $433~\rm{MHz}$ &  $100~\rm{bps}$ &Very Low & Long\\ \hline

RedCap  &   $20~\rm{MHz}$  & $85~\rm{Mbps}-225~\rm{Mbps}$ &  High & Medium\\ \hline

eRedCap &    $1.5~\rm{GHz}-24~\rm{GHz}$   &  $10~\rm{Mbps}$   & Medium & Medium\\ \hline

LoRa  &     $433~\rm{MHz}$, $868~\rm{MHz}$, $915~\rm{MHz}$  &   $0.5~\rm{kbps}-50~\rm{kbps}$  & Low & Long \\ \hline

\end{tabular}
\end{table}

The LoRa Alliance has been crucial in  leading to its commercialization and promoting the implementation of LoRa across various fields, such as manufacturing \cite{ref35,ref36}, IoT positioning \cite{ref37,ref86}, and satellite communication \cite{ref38,ref39, ref87}.\footnote{https://lora-alliance.org} Due to the patent protection of the LoRa physical layer \cite{ref47}, early studies primarily concentrated on experimental investigations, aiming to evaluate various performance metrics of LoRa modulation. In \cite{ref41}, a stochastic geometry framework has been provided to model the performance of a single gateway LoRa network and enables the rigorous scalability analysis of LoRa network. A sub-optimal SF allocation scheme is proposed in \cite{ref42} to maximize average system packet success probability by properly allocating SF to each traffic, which also maximizes the connectivity of the end devices. In addition, the theoretical maximum number of end devices of a single LoRaWAN gateway has been investigated, and its performance at the physical and data link layers has been evaluated in \cite{ref43, ref44}, respectively.

\begin{table*}[]
\centering
\label{tab1}
\caption{\textsc{Main Notations Used in This Paper.}}
\begin{tabular}{|m{1cm}<{\centering}|m{7cm}<{\centering}|m{1cm}<{\centering}|m{7.3cm}<{\centering}|}

\hline
Notation & Indication &  Notation & Indication     \\ \hline
$f$  & The SF of the LoRa signal& $E_\mathrm{s}$     &      The symbol energy \\ \hline
$T_{\mathrm{chip}}$   &     The chip duration    &  $f_\mathrm{s}$  & The start frequency of the LoRa signal   \\ \hline
$B_\mathrm{w}$   &   The bandwidth of the LoRa signal    & $N_{\mathrm{com}}$  & The number of SFs combinations  \\ \hline
$M$   &   The number of chirp signal types in the transmitted signal  & $N_{\mathrm{av}}$ &   The number of available types for the chirp signal \\ \hline
$x_{i,q}(n)$ & The chirp signal of the $q$-th sub-block of the $i$-th block & $x(n)$ & The transmitted signal of the proposed SFI-LoRa scheme \\ \hline
$\textbf{S}_\delta$  & The $\delta$-th SF combination in $\textbf{S}$  & $\textbf{S}_\mathrm{av}$&The set of all available SFs  \\ \hline
$\textbf{S}_\mathrm{ns}$&The set of the remaining $N_{\mathrm{av}}-M$ SFs &$\textbf{S}_{\mathrm{pre}}$  & The  sequence getting by combinatorial method\\ \hline
$\bar{\omega}_{f,m}$  & The basis function of $w_o(nT_{\mathrm{chip}})$ with the SF $f$ and the modulated symbol $m$ & $w_o$  & The $o$-th transmitted signal of conventional LoRa \\ \hline
$h_\mathrm{b}$  & The transmitting antenna height & $\textbf{D}$  & The set of all modulated symbols during a symbol duration\\ \hline
$h_\mathrm{r}$  & The receiving antenna height & $a(h_\mathrm{r})$  & The correction factor for the mobile station antenna height $h_\mathrm{r}$\\ \hline
$N_{\mathrm{tot}}$  &  The number of information bits transmitted during a symbol duration in the proposed SFI-LoRa scheme &$N_{\mathrm{sub},i}$ & The number of sub-blocks in $i$-th block\\ \hline
$\Theta$  &  The data rate of the proposed SFI-LoRa scheme & $P_{\mathrm{pe}|i,j,h}$ & The probability of error detection given that the $i$-th element of \(\textbf{S}_\mathrm{est}\) and the $j$-th element of  \(\textbf{S}_\mathrm{ns}\)\\ \hline
\end{tabular}
\end{table*}

Nevertheless, LoRa utilizes CSS modulation to attain ultra-low receiver sensitivity. This approach inherently results in a trade-off between the data rate and coverage range.
As shown in Table \ref{tab0}, the data rate of LoRa is significantly lower than other popular wireless technologies such as Wi-Fi, Bluetooth, and ZigBee. To address issue, several studies have contributed to improve the data rate of LoRa networks. In \cite{ref89}, a multiple-input multiple-output (MIMO)-aided LoRa scheme has been developed for achieving high data rate transmissions. Efficient coherent and noncoherent demodulation algorithms for the MIMO-LoRa scheme have been proposed by leveraging properties of modulated chirps in \cite{ref88}. In \cite{ref92}, novel joint delta scripting algorithm and concatenated channel coding scheme have been proposed to resolve the link quality indeterminacy. A deep-neural-network-based LoRa (NELoRa) demodulation scheme has been proposed in \cite{ref60}. The NELoRa which lowers the signal-to-noise ratio (SNR) threshold of the chirp symbol decoding at a single gateway can improving the overall network throughput. In \cite{ref61}, frequency hopped chirp spread spectrum (FHCSS) has been proposed to recover collided packets. The efficient carrier-sense-multiple-access-based LoRa media access control (LMAC) proposed in \cite{ref62} implements carrier-sense multiple access. The LMAC balances the communication loads among the channels defined by frequencies and spreading factors based on the end nodes local information and the gateways global information. In \cite{ref63}, $p$ persistent-channel activity recognition multiple access ($p$-CARMA) has been proposed to reduce collisions. The $p$-CARMA enables higher packet reception ratio as well as a higher number of received packets by utilizing the LoRaWAN channel activity detection. Although these proposals enhance channel utilization by addressing collisions, designing MAC layer and detecting channel activity, the single-link throughput still remain restricted.
\begin{figure*}[t]
    \centering
    \includegraphics[width=1\linewidth]{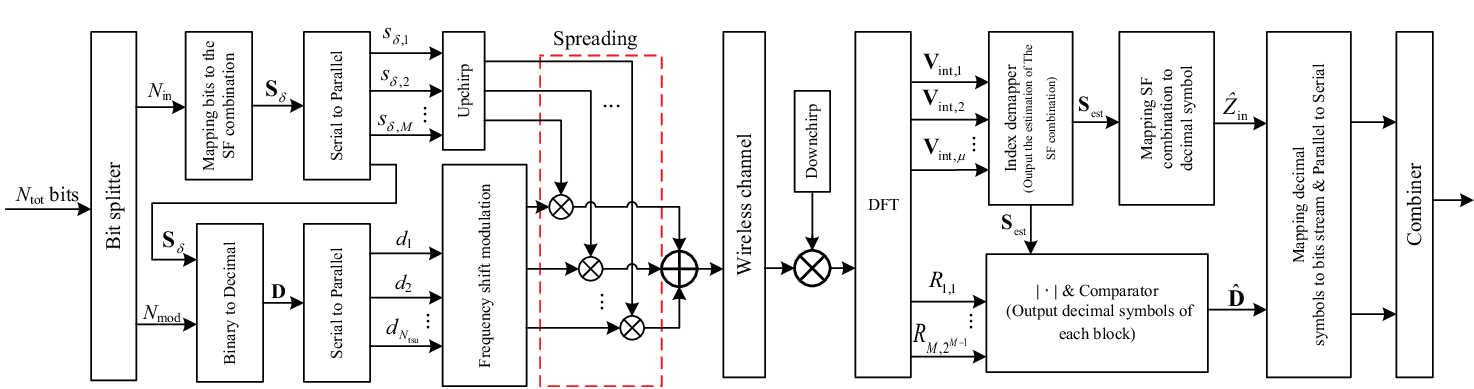}
    \caption{System model of the proposed SFI-LoRa scheme.}
    \label{fig1}
\end{figure*}

As a result, some researchers have attempted to increase the throughput of individual links through physical layer modifications. In \cite{ref23}, an interleaved chirp spreading LoRa (ICS-LoRa) has been proposed to split the LoRa symbol into smaller parts for interleaving transmission to increase the data rate. A slope-shift keying LoRa (SSK-LoRa) has been proposed in \cite{ref24}. The SSK-LoRa employs a linearly varying-frequency down-chirp and its cyclic shifts to obtain a new signal dimension,  which can be utilized to carry an additional information bits. In \cite{ref9}, phase-shift keying (PSK) modulation is used in PSK-LoRa to convey additional information bits. However, PSK-LoRa requires a coherent receiver, which increases both the cost and complexity. A frequency-bin-index LoRa (FBI-LoRa) has been proposed in \cite{ref11}, which employs index modulation (IM) to improve the data rate. Unfortunately, the bit error rate (BER) performance of FBI-LoRa is worse than the conventional LoRa. In \cite{ref10}, a multiplexed LoRa (MuLoRa) has been proposed to improve the data rate and spectral efficiency. Nevertheless, the MuLoRa scheme has similar BER performance to the conventional LoRa scheme. In the MuLoRa scheme, fractional Fourier transform (FRFT) is used for demodulation, which leads to relatively high computational complexity. A hybrid modulation LoRa (HM-LoRa) has been proposed in \cite{ref93}, which utilizes the quasi-orthogonality of the LoRa signals with different SFs to improve data rate. However, these schemes either slightly increase data rate or sacrifice BER performance. Hence, a better trade-off between data rate and BER performance still deserves further exploration.

Based on the motivation outlined above, we propose a novel spreading-factor-index (SFI)-based LoRa scheme to achieve high-data-rate transmissions. The main contributions of this paper are summarized as follows.

1) We propose a novel LoRa modulation scheme, namely SFI-LoRa modulation, and put forward its transceiver architecture. In the proposed SFI-LoRa scheme, we establish the combinations of SFs as a new set of indices, which is treated as a new signal dimension to convey additional information bits.

2) We conduct theoretical derivation on the symbol error rate (SER) of the proposed SFI-LoRa modulation scheme over additive white Gaussian noise (AWGN) and Rayleigh fading channels. The accuracy of the theoretical SER expressions is validated by simulations under various conditions.

3) We carefully analyze the data rate, transmission throughput, complexity and energy efficiency of proposed SFI-LoRa modulation scheme and compare these performance metrics with the state-of-the-art counterparts.

4) We perform a variety of simulations to demonstrate that the proposed scheme can achieve desirable throughput gains over the existing LoRa schemes  without sacrificing the BER performance over AWGN, Rayleigh fading, and multipath flat-fading channels. We also discuss the coverage probability of our design to illustrate its feasibility in practical applications.
In particular, it is observed that the proposed scheme has throughput gains of 170\% (resp. 211\% and 168\%), 52\% (resp. 75\% and 51\%), 35\% (resp.  56\% and 34\%), 14\% (resp. 12\% and 7\%) compared to the conventional LoRa, MuLoRa, DCDSK, and HM-LoRa schemes, respectively, over a AWGN channel (resp. Rayleigh fading channel and multipath flat-fading channel). Furthermore, our scheme achieves a BER performance gain of 28.8\% (resp. 34.9\% and 62.1\%) compared to the FBI-LoRa scheme with a slight throughput performance loss of $14.8\%$ (resp. 16\% and 5\%) over AWGN channel (resp. Rayleigh fading channel and multipath flat-fading channel).

The structure of this paper is organized as follows. Section II provides fundamental background of LoRa modulation. Section III illustrates the system model of the proposed SFI-LoRa scheme. Theoretical analyses regarding the SER performance are derived in Section IV. Section V presents various numerical results and discussions. Finally, Section VI concludes this paper. To facilitate the exposition of proposed scheme, Table II summarizes critical mathematical notations used in this paper.

\section{Fundamentals of LoRa Modulation}
The frequency variation of the LoRa signal is increasing linearly with time. A LoRa symbol is divided into $2^f$ chips, where the SF $f\in\{7,8,...,12\}$ \cite{ref33, ref14}. Therefore, the chip period of LoRa is $T_{\mathrm{chip}}=\frac{1}{B_\mathrm{w}}$ and the symbol period is $T_\mathrm{sym}=2^f\cdot T_{\mathrm{chip}}$.  Since each LoRa symbol transmits $f$ bits, the bit rate of the conventional LoRa scheme can be expressed as  $R_{\mathrm{b}}=\frac{f}{2^{f}}\cdot B_{\mathrm{w}}\cdot CR$, where $CR$ denotes coding rate. We assume the $o$-th transmitted symbol is $d_o=m\in\{0,1,...,2^f-1\}$. The frequency of the LoRa symbol $d_o$ rises from the initial frequency $f_\mathrm{s}=\frac{B_\mathrm{w}\cdot m}{2^f}$ to $B_\mathrm{w}$, and then returns to 0 after a jump at $T_0=(2^f-m)/B_\mathrm{w}$. During the remaining symbol duration, the frequency
rises to $f_\mathrm{s}$ again \cite{ref33}. Therefore, the discrete-time LoRa signal expression of transmitted symbol $d_o$ is \cite{ref33}
\begin{equation}\label{f1}\begin{aligned}
    w_o(nT_{\mathrm{chip}})&=\sqrt{E_\mathrm{s}}\cdot\bar{\omega}_{f,m}(nT_{\mathrm{chip}})\\
    &=\sqrt{\frac{E_\mathrm{s}}{2^{f}}}\exp\left[j2\pi\left(\frac{\left((m+n)\mathrm{mod}2^{f}\right)^2}{2^{f+1}}\right)\right],
\end{aligned}
\end{equation}
where $n$ is the index at time $nT_{\mathrm{chip}}$, $E_\mathrm{s}$ is the symbol energy, and $\bar{\omega}_{f,m}(nT_{\mathrm{chip}})$ is the basis function of $w_o(nT_{\mathrm{chip}})$. The (\ref{f1}) illustrates that the transmitted LoRa signal can be readily obtained via a discrete circular time shift operation. Since the chirp signals with distinct offsets are orthogonal to each other. The correlation between the transmitted signal $w_o(nT_{\mathrm{chip}})$ and $2^f$ possible LoRa signals satisfies the following property \cite{ref14}
\begin{equation}\Omega_\lambda=\sum_{n=0}^{2^{f}-1}w_o\left(nT_{\mathrm{chip}}\right)\cdot\omega_{f,\lambda}^*\left(nT_{\mathrm{chip}}\right)=
\begin{cases}
\sqrt{E_\mathrm{s}}, & \lambda=m \\
0, & \lambda \neq m
\end{cases}.\end{equation}
As a result, the above property can be used to demodulate the LoRa signal. First, the received signal is multiplied with the conjugated raw
up-chirp $\omega_{f,0}^*(nT_{\mathrm{chip}})$ \cite{ref33}, hereafter named down-chirp, where $\omega_{f,0}^*(nT_{\mathrm{chip}})$ can be expressed as
\begin{equation}\omega_{f,0}^*\left(nT_{\mathrm{chip}}\right)=\sqrt{\frac{1}{2^{f}}}\exp\left(-j2\pi\frac{n^{2}}{2^{f+1}}\right).\end{equation}
Subsequently, the calculation process of performing discrete Fourier transform (DFT) on the dechirped signal can be obtained as
\begin{equation}\dot{\Omega}_{\mathrm{DFT}}=\mathrm{DFT}\left(\mathbf{r}_{o}\circ\mathbf{w}_{f,0}^*\right),\end{equation}
where $\mathbf{r}_{o}=[r_o(T_{\mathrm{chip}}),r_o(2T_{\mathrm{chip}}),...,r_o(2^fT_{\mathrm{chip}})]$, $\mathbf{w}_{f,0}^*=[\omega_{f,0}^*(T_{\mathrm{chip}}),\omega_{f,0}^*(2T_{\mathrm{chip}}),...,\omega_{f,0}^*(2^fT_{\mathrm{chip}})]$, and $\circ$ is the Hadamard product operator.
In the following step, the transmitted LoRa symbol is recovered by selecting the frequency bin index via the peak detection, which is given by
\begin{equation}\hat{d}_o=\arg\max_{i=0,\ldots,2^{f}-1}\left(\left|\dot{\Omega}_{\mathrm{DFT},\lambda}\right|\right).\end{equation}

\section{System Model of Proposed SFI-LoRa Scheme}

In this section, we elaborate on the transceiver architecture of the proposed SFI-LoRa scheme, as well as illustrate its transmission mechanism.

\subsection{Transmitter of the Proposed SFI-LoRa}
The components of the proposed SFI-LoRa scheme is shown in Fig. \ref{fig1}. In our design, \(N_{\mathrm{tot}}=N_{\mathrm{in}}+N_{\mathrm{mod}}\) information bits are transmitted, consisting of \(N_{\mathrm{in}}\) index bits and \(N_{\mathrm{mod}}\) modulated bits. The index bits are conveyed by the SF combinations, and modulated bits are transmitted by the starting frequency bin (SFB) of chirp signal like the conventional LoRa. In other words, the proposed scheme utilizes the indices of different SF combinations as a new dimension to transmit additional information bits.

At the transmitter, \(M\) types of chirp signals are selected from the \(N_\mathrm{av}\) available chirp-signal types as a SF combination to convey the index bits $\tau_\mathrm{in}$, where \(M\in\{1,2,...,N_\mathrm{av}-1\}\).\footnote{The variable \(N_{\mathrm{av}}\) is the number of available types for the chirp signal. According to \cite{ref33},  \(N_{\mathrm{av}}=6\) is considered in this paper. Specifically, the available SFs are denoted as \(\textbf{S}_{\mathrm{av}}=\{s_{\mathrm{av,1}},...,s_{\mathrm{av,\mu}},...s_{\mathrm{av,}N_\mathrm{av}}\}=\{7,...,12\}\).} It can be observed that the number of SF combinations is $N_{\mathrm{com}}=\left(\begin{smallmatrix}N_{\mathrm{av}}\\M\end{smallmatrix}\right)$, where
 \begin{equation}
     \begin{pmatrix}X\\Y\end{pmatrix}=\frac{X!}{Y!(X-Y)!}.
 \end{equation}
Thus, the number of index bits can be calculated by  \begin{equation}\label{f7}
N_{\mathrm{in}}=\left\lfloor\log_2\begin{pmatrix}N_{\mathrm{av}}\\M\end{pmatrix}\right\rfloor,
\end{equation}
where$\lfloor\cdot\rfloor$ denotes the floor operation. Here, we define
\begin{equation}\mathbf{\textbf{S}}=
\begin{bmatrix}
\textbf{S}_1 \\
\vdots\\
\textbf{S}_\delta\\
\vdots\\
\textbf{S}_{N_{\mathrm{com}}}
\end{bmatrix}=
\begin{bmatrix}
s_{1,1}&\cdots&s_{1,i}&\cdots&s_{1,M} \\
\vdots&\ddots&\vdots&\ddots&\vdots\\
s_{\delta,1}&\cdots&s_{\delta,i}&\cdots&s_{\delta,M}\\
\vdots&\ddots&\vdots&\ddots&\vdots\\
s_{N_{\mathrm{com}},1}&\cdots&s_{N_{\mathrm{com},i},i}&\cdots&s_{N_{\mathrm{com}},M}
\end{bmatrix}
\end{equation}
as a set of different SF combinations, where \(1\leq\delta\leq N_{\mathrm{com}}\). The $\delta$-th SF combination is defined as \(\textbf{S}_\delta=\{s_{\delta,1},...,s_{\delta,i},...,s_{\delta,M}\}\), and the element $s_{\delta,i}$ of $\textbf{S}_\delta$ represents the selected SF. It should be emphasized that the selected SF \(s_{\delta,i}\) is unique in $\textbf{S}_\delta$ and the elements of SF combination \(\textbf{S}_\delta\) is strictly monotonically decreasing from \(i=1\) to \(i=M\). Moreover, the index symbol \(Z_\mathrm{in}\) of index bits $\tau_\mathrm{in}$ is mapped to a fixed SF combination using combinatorial method.

We now provide a brief introduction to the combinatorial method. Specifically, for fixed values of \(N_{\mathrm{av}}\) and \(M\), the index symbol \(Z_\mathrm{in}\in \left[0,\left(\begin{smallmatrix}N_{\mathrm{av}}\\M\end{smallmatrix}\right)-1\right]\) can be changed into a mapped sequence \(\textbf{S}_\mathrm{pre}=\{s_{\mathrm{pre},1},...,s_{\mathrm{pre},M}\}\), where \(s_{\mathrm{pre},1}>\cdots>s_{\mathrm{pre},M}\geq0\) and \(s_{\mathrm{pre},1},\cdots,s_{\mathrm{pre},M} \in \{0,...,N_{\mathrm{av}}-1\}\). The elements of \(\textbf{S}_\mathrm{pre}\) can be obtained as
\begin{equation}\label{f8}
Z_\mathrm{in}=\begin{pmatrix}s_{\mathrm{pre},1}\\M\end{pmatrix}+\cdots+\begin{pmatrix}s_{\mathrm{pre},M-1}\\2\end{pmatrix}+\begin{pmatrix}s_{\mathrm{pre},M}\\1\end{pmatrix}.
\end{equation}
For all $Z_\mathrm{in}$ and \(M\), the method for generating \(s_{\mathrm{pre},i}\)
is described as follows: Initially, the procedure determines the maximum value \(s_{\mathrm{pre},1}\) that satisfies \(\left(\begin{smallmatrix}s_{\mathrm{pre},1}\\M\end{smallmatrix}\right) \leq Z_\mathrm{in}\). Subsequently, the procedure identifies the largest \(s_{\mathrm{pre},2}\) that satisfies \(\left(\begin{smallmatrix}s_{\mathrm{pre},1}\\M\end{smallmatrix}\right)+\left(\begin{smallmatrix}s_{\mathrm{pre},2}\\M-1\end{smallmatrix}\right) \leq Z_\mathrm{in}\) and continues this process until the mapped sequence \(\textbf{S}_\mathrm{pre}\) is generated. In the proposed SFI-LoRa scheme, the index bits $\tau_\mathrm{in}$ are first converted to the index symbol \(Z_\mathrm{in}\) at the transmitter, then \(Z_\mathrm{in}\) is transformed into the mapped sequence \(\textbf{S}_\mathrm{pre}\) by the combinatorial method. The values in \(\textbf{S}_\mathrm{pre}\) are constrained to the range from \(0\) to \(N_\mathrm{av}-1\). Subsequently, note that the available SFs $\textbf{S}_\mathrm{av}$ and the SF combination \(\textbf{S}_\mathrm{\delta}\) have the same number of elements and a one-to-one correspondence,  the elements of \(\textbf{S}_\mathrm{\delta}\) can be derived using the linear transformation \(s_{\mathrm{\delta},i}=s_{\mathrm{av},i}+7\).

Moreover, the corresponding transmitted signal is generated by the SF combination \(\textbf{S}_\delta\) to convey the \(N_{\mathrm{mod}}\) modulated bits. Specifically, the transmitted signal is superposed of \(M\) blocks, and the $i$-th block of the transmitted signal is modulated by the SF \(s_{\delta,i}\) (\(i\in\{1,...,M\}\)). Furthermore, \(N_{\mathrm{sub},i}\cdot s_{\delta,i}\) modulated bits are transmitted in the $i$-th block, where \(N_{\mathrm{sub},i}=2^{i-1}\) means the number of sub-blocks in the $i$-th block. The symbol duration of the LoRa signal is related to the SF and can be denoted as $T_{\mathrm{sym}}=2^f\cdot T_{\mathrm{chip}}$. To improve the spectrum efficiency, each block is composed of different numbers of the same type of LoRa signals. In other words, the \(i\)-th block can be split into \(N_{\mathrm{sub},i}\) sub-blocks. The $q$-th sub-block is modulated by the SF $s_{\delta,i}$ and transmits modulated bits \(\tau_{\kappa}\) by the SFB, where $\kappa=2^{i-1}+(q-1)$ and $q\in \{1,...,N_{\mathrm{sub},i}\}$. Notably, $N_{\mathrm{tsu}}=\sum_{i=1}^{M}N_{\mathrm{sub},i}$ modulated symbols are transmitted through the SFB within a symbol duration. We define these information bits and modulated symbols as $\{\tau_1,...,\tau_{\kappa},...,\tau_{N_{\mathrm{tsu}}}\}$ and $\textbf{D}=\{d_1,...,d_{\kappa},...,d_{N_{\mathrm{tsu}}}\}$, respectively. As a result, the $q$-th sub-block transmitted signal of the $i$-th block can be represented as
\begin{equation}
\begin{split}\label{f10}
x_{i,q}(nT_{\mathrm{chip}})\!=&\!\sqrt{\frac{E_\mathrm{s}}{M\!\cdot\! 2^{i-1}\!\cdot \! 2^{s_{\delta,i}}}}\!\!\\&\times\exp\biggl[j2\pi\!\cdot\!\frac{((d_\kappa+n)\mathrm{mod}2^{s_{\delta,i}})^2}{2^{s_{\delta,i}+1}}\biggr],
\end{split}
\end{equation}
where \(E_\mathrm{s}\) denotes the symbol energy, \(n\) denotes the index of the sample at time \(nT_{\mathrm{chip}}\).

\begin{figure}[t]
    \centering
    \includegraphics[width=1\linewidth]{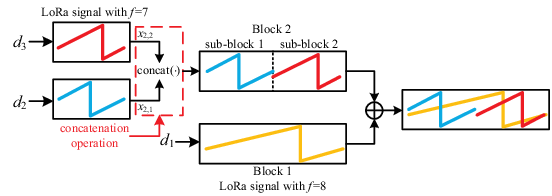}
    \caption{The transmitted signal of the proposed SFI-LoRa scheme.}
    \label{fig2}
\end{figure}

To illustrate a little further, a case of converting index bits $\tau_\mathrm{in}$ to \(\textbf{S}_\mathrm{\delta}\) is shown in Table  \ref{tab2}. For example, when $M=2$ and the index bits $\tau_\mathrm{in}=000$, the SF combination can be obtained by (\ref{f8}) and represented as $\textbf{S}_1=\{s_{1,1},s_{1,2}\}=\{8,7\}$. Moreover, Fig.~\ref{fig2} presents the transmitted signal of the proposed SFI-LoRa scheme under the SF combination $\textbf{S}_1$. One can observe that Block 2 consists of two LoRa signals with the SF $s_{1,2}=7$ through the concatenation operation $\mathrm{concat}(\cdot)$, where the operation $\mathrm{concat}(x_{2,1},x_{2,2})$ can be expressed by \begin{equation}\label{f11}
    \mathrm{concat}(x_{2,1},x_{2,2})=[x_{2,1}(nT_{\mathrm{chip}}),x_{2,2}(nT_{\mathrm{chip}})].
\end{equation}
The transmitted signal is then formed by superposing the modulated LoRa signals across all \(M\) blocks. Eventually, as shown in
Fig. \ref{fig1}, the transmitted signal of the proposed SFI-LoRa scheme can be expressed as
\begin{equation}
\label{f3}
\begin{split}
x(nT_{\mathrm{chip}})& =\sum_{i=1}^M\mathrm{concat}(x_{i,1},x_{i,2},...,x_{i,N_{\mathrm{sub},i}}).
\end{split}
\end{equation}
As a further illustration, the details of the proposed SFI-LoRa modulation process is summarized in Algorithm \ref{modulate}.

\begin{table}[t]
\centering
\renewcommand\arraystretch{1.2}
\caption{\textsc{An Example of Combinatorial Method}}
\label{tab2}
\begin{tabular}{|c|c|c|c|c|}

\hline
$\tau_\mathrm{in}$ & $Z_\mathrm{in}$ & Combinatorial Method   & $\textbf{S}_{\mathrm{pre}}$ & $\textbf{S}_{\delta}$   \\ \hline

$111$ & $7$ & $\begin{pmatrix}4\\2\end{pmatrix}+\begin{pmatrix}1\\1\end{pmatrix}$   & $\{4,1\}$ & $\{11,8\}$   \\ \hline

$110$ & $6$ & $\begin{pmatrix}4\\2\end{pmatrix}+\begin{pmatrix}0\\1\end{pmatrix}$   & $\{4,0\}$ & $\{11,7\}$   \\ \hline

$\vdots$ &$ \vdots$ & $\vdots$   & $\vdots$ & $\vdots$ \\ \hline

$001$ & $1$ & $\begin{pmatrix}2\\2\end{pmatrix}+\begin{pmatrix}0\\1\end{pmatrix}$   & $\{2,0\}$ & $\{9,7\}$   \\ \hline

$000$ & $0$ & $\begin{pmatrix}1\\2\end{pmatrix}+\begin{pmatrix}0\\1\end{pmatrix}$   & $\{1,0\}$ & $\{8,7\}$   \\ \hline

\end{tabular}
\end{table}

\setlength{\textfloatsep}{10pt}


\begin{algorithm}[t]
    \caption{Modulation algorithm for the proposed SFI-LoRa scheme}
    \label{modulate}

    \begin{algorithmic}[1]
        \State  \textbf{Input:} $M$, $N_\mathrm{av}$, and $\textbf{S}_\mathrm{av}$.
        \State \textbf{Output:} The transmitted signal $x(nT_{\mathrm{chip}})$.
        \State  \textbf{Step 1:} Generate the SF combination
        \State  $N_{\mathrm{in}}=\left\lfloor\log_2\begin{pmatrix}N_{\mathrm{av}}\\M\end{pmatrix}\right\rfloor$;
        \State  Map index bits $\tau_\mathrm{in}$ to a fixed SF combination $\textbf{S}_\delta=\{s_{\delta,1},...,s_{\delta,i},...,s_{\delta,M}\}$ by the combinatorial method;
        \State  \textbf{Step 2:} Generate LoRa signals
        \For{$i=1:1:M$}
            \State $N_{\mathrm{sub},i}=2^{i-1}$;
            \For{$q=1:1:N_{\mathrm{sub},i}$}
                \State $\tau_\kappa \to d_{\kappa}$, where $\kappa=2^{i-1}+(q-1)$;
                \State Generate the LoRa signal $x_{i,q}(nT_{\mathrm{chip}})$ through the
                \Statex \quad\quad\quad transmitted symbol $d_{\kappa}$ by employing (\ref{f10});
            \EndFor
        \EndFor
        \State  \textbf{Step 3:} Generate the transmitted signal
        \For{$i=1:1:M$}
            \State Concatenate all sub-blocks of the $i$-th block by em-
            \Statex \hspace*{1.1em} ploying (\ref{f11});
        \EndFor
        \State Superpose all blocks to generate the transmitted signal $x(nT_{\mathrm{chip}})$.

    \end{algorithmic}
\end{algorithm}

\subsection{Receiver of the Proposed SFI-LoRa}
In the proposed SFI-LoRa scheme, the received signal can be given by
\begin{equation}
    r(nT_{\mathrm{chip}})=\sqrt{h}x(nT_{\mathrm{chip}})+z(nT_{\mathrm{chip}}),
\end{equation}
where \(\sqrt{h}\) represents the magnitude of fading channel coefficient, \(z(nT_{\mathrm{chip}})\) is the complex AWGN. Similarly, the elements of the estimated SF combination \(\textbf{S}_\mathrm{est}\) are strictly monotonically decreasing from \(i=1\) to \(i=M\). As mentioned above, the symbol duration of a LoRa signal is related to the SF. Therefore, we intercept the received signal according to the $\mu$-th available SF $s_{\mathrm{av},\mu}$ during the traversal of the available SFs $\textbf{S}_\mathrm{av}$. In other words, the received signal from $T_{\mathrm{chip}}$ to $2^{s_{\mathrm{av,\mu}}}T_\mathrm{chip}$ is first extracted based on the SF $s_{\mathrm{av,\mu}}$, which can be formulated as
\begin{equation}
    \textbf{r}_{\mathrm{int},\mu}=[r(1T_{\mathrm{chip}}),r(2T_{\mathrm{chip}}),...,r(2^{s_{\mathrm{av,\mu}}}T_\mathrm{chip})].
\end{equation}
The $ \textbf{r}_{\mathrm{int},\mu}$ is then dechirped and processed by the DFT to obtain
\begin{equation}\label{f5}\begin{aligned}
\textbf{V}_{\mathrm{int},\mu}=\bigl|\mathrm{DFT\bigl(} \textbf{r}_{\mathrm{int},\mu}\circ\textbf{w
}_{s_{\mathrm{av,\mu}},0}^*\bigr)\bigr|, \end{aligned}\end{equation}
where \(\textbf{V}_{\mathrm{int},\mu}=[v_{\mathrm{int},1},v_{\mathrm{int},2},...,v_{\mathrm{int},2^{s_{\mathrm{av,\mu}}}}]\) and \(\textbf{w
}_{s_{\mathrm{av,\mu}},0}^*\!=\![w_{s_{\mathrm{av,\mu}},0}^*(1T_\mathrm{chip})\!,\!w_{s_{\mathrm{av,\mu}},0}^*(2T_\mathrm{chip})\!,\!...,\!w_{s_{\mathrm{av,\mu}},0}^*(2^{s_{\mathrm{av,\mu}}}T_\mathrm{chip})]\). Afterwards, the receiver gets the maximum amplitude of the \(\textbf{V}_{\mathrm{int},\mu}\). Repeat the above steps until all elements in $\textbf{S}_\mathrm{av}$ have been traversed and $N_\mathrm{av}$ maximum amplitudes are generated. These maximum amplitudes can be represented by \begin{equation}
    \textbf{c}=\{c_{1},...,c_{\mu},...,c_{N_{\mathrm{av}}}\}.
\end{equation} Subsequently, the mapped sequence $\hat{\textbf{S}}_{\mathrm{pre}}$ can be recovered by selecting the index of the $M$ maximum amplitudes in $\textbf{c}$, as
\begin{equation}
\begin{split}
\hat{\textbf{S}}_{\mathrm{pre}}&=\arg\max^M(\textbf{c})\\&
=\{\hat{s}_{\mathrm{pre},1},...,\hat{s}_{\mathrm{pre},i},...,\hat{s}_{\mathrm{pre},M}\},
\end{split}
\end{equation}
where $\arg\max\limits^M(\textbf{A})$ denotes the output index corresponding to the $M$ largest values in the set $\textbf{A}$. The elements of \(\textbf{S}_\mathrm{est}\) can be derived using the linear transformation \(s_{\mathrm{est},i}=\hat{s}_{\mathrm{pre}}+7\). Thus, the index symbol $\hat{Z}_\mathrm{in}$ and index bits $\hat{\tau}_\mathrm{in}$ can be obtained by the combinatorial method.

\begin{algorithm}[t]
    \caption{Demodulation algorithm for the proposed SFI-LoRa scheme}
    \label{demodulate}
    \begin{algorithmic}[1]
        \State  \textbf{Input:} $M$, $N_\mathrm{av}$, $\textbf{S}_\mathrm{av}$, $N_{\mathrm{tsu}}$, and $r(nT_{\mathrm{chip}})$.
        \State  \textbf{Output:} The index bits $\hat{\tau}_\mathrm{in}$ and modulated bits $\{\hat{\tau}_1,...,\hat{\tau}_{\kappa},...,\hat{\tau}_{N_{\mathrm{tsu}}}\}$.
        \State  \textbf{Step 1:} Recover the index bits

        \For{$\mu=1:1:N_{\mathrm{av}}$}
                \State $\textbf{r}_{\mathrm{int},\mu}=[r(1T_{\mathrm{chip}}),r(2T_{\mathrm{chip}}),...,r(2^{s_{\mathrm{av,\mu}}}T_\mathrm{chip})]$;
                \State $\textbf{V}_{\mathrm{int},\mu}=\bigl|\mathrm{DFT\bigl(} \textbf{r}_{\mathrm{int},\mu}\circ\textbf{w}_{s_{\mathrm{av,\mu}},0}^*\bigr)\bigr|$;
                \State $c_{\mu}=\mathrm{max}(\textbf{V}_{\mathrm{int},\mu})$;
        \EndFor
        \State $\hat{\textbf{S}}_{\mathrm{pre}}=\arg\max\limits^M(\textbf{c})$;
        \For{$i=1:1:M$}
                \State \(s_{\mathrm{est},i}=\hat{s}_{\mathrm{pre},i}+7\);
        \EndFor
        \State Convert $\textbf{S}_{\mathrm{est}}$ to $\hat{Z}_\mathrm{in}$ by the combinatorial method;
        \State $\hat{Z}_\mathrm{in} \to \hat{\tau}_\mathrm{in}$;
        \State  \textbf{Step 2:} Recover the modulated bits
        \For{$i=1:1:M$}
            \For{$\eta=1:1:N_{\mathrm{sub},i}$}
                \State $\textbf{r}_{\mathrm{sub},i,\eta}=[r((1+(\eta-1)2^{s_{\mathrm{est},i}})T_{\mathrm{chip}}),r((2+(\eta-$
                \Statex \quad\quad\quad $1)2^{s_{\mathrm{est},i}})T_{\mathrm{chip}}),...,r(\eta\cdot2^{s_{\mathrm{est},i}}T_\mathrm{chip})]$;
                \State $R_{i,\eta}=\bigl|\mathrm{DFT}\bigl( \textbf{r}_{\mathrm{sub},i,\eta}\circ\textbf{w}_{s_{\mathrm{est},i},0}^*\bigr)\bigr|$;
                \State $\hat{d}_{\kappa}=\arg\max_{\eta\in\{1,...,N_{\mathrm{sub},i}\}}\left(|R_{i,\eta}|\right)$, where $\kappa=$
                \Statex \quad\quad\quad $2^{i-1}+(\eta-1)$;
            \EndFor
        \EndFor
        \State $\{\!\hat{d}_1\!,\!...,\!\hat{d}_{\kappa}\!,\!...,\!\hat{d}_{N_{\mathrm{tsu}}}\} \!\to \! \{\!\hat{\tau}_1\!,\!...,\!\hat{\tau}_{\kappa}\!,\!...,\!\hat{\tau}_{N_{\mathrm{tsu}}}\!\}$.

    \end{algorithmic}
\end{algorithm}

Furthermore,  we traverse the $\textbf{S}_\mathrm{est}$ to recover the modulated bits. During the traversal process based on the $\textbf{S}_{\mathrm{est}}$, the received signal consists of $N_{\mathrm{sub},i}$ components and is given by
\begin{equation}
    r(nT_\mathrm{chip})=\mathrm{concat}(\textbf{r}_{\mathrm{sub},i,1},...,\textbf{r}_{\mathrm{sub},i,\eta},...,\textbf{r}_{\mathrm{sub},i,\hat{N}_{\mathrm{sub},i}}),
\end{equation}
where $\textbf{r}_{\mathrm{sub},i,\eta}=[r((1+(\eta-1)2^{s_{\mathrm{est},i}})T_{\mathrm{chip}}),r((2+(\eta-1)2^{s_{\mathrm{est},i}})T_{\mathrm{chip}}),...,r(\eta\cdot2^{s_{\mathrm{est},i}}T_\mathrm{chip})],$
and $\eta\in\{1,...,N_{\mathrm{sub},i}\}$.
The modulated symbol of $\textbf{r}_{\mathrm{sub},i,\eta}$ can be yielded as \begin{equation}\begin{aligned}\hat{d}_{\kappa}=\arg\max_{\eta\in\{1,...,N_{\mathrm{sub},i}\}}\left(|R_{i,\eta}
|\right)\},\end{aligned}\end{equation}
where
\begin{equation}\label{f22}
    \begin{aligned}
        R_{i,\eta}&=\frac{1}{\sqrt{2^{s_{\mathrm{est},i}}}}\sum_{k=0}^{2^{s_{\mathrm{est},i}}-1}
        \tilde{\textbf{r}}_{i,\eta}e^{-j2\pi\frac{k\lambda}{2^{s_{\mathrm{est},i}}}} \\& =
        \begin{cases}
        \sqrt{\frac{h E_{\mathrm{s}}}{M\cdot 2^{i-1}\cdot 2^{s_{\mathrm{est},i}}}}+\phi_{s_{\mathrm{est},i}}, & \quad\lambda=d_{t}, s_{\mathrm{est},i}\in \textbf{S}_{\delta} \\
        \phi_{s_{\mathrm{est},i}}, & \quad\lambda\neq d_{t}, s_{\mathrm{est},i}\in \textbf{S}_{\delta} \\
        \phi_{s_{\mathrm{est},i}}, & \quad s_{\mathrm{est},i}\notin \textbf{S}_{\delta} \\
        \end{cases}.
    \end{aligned}
\end{equation}
In (\ref{f22}), $\tilde{\textbf{r}}_{i,\eta}=\textbf{r}_{\mathrm{sub},i,\eta}\circ\textbf{w}_{s_{\mathrm{est},i},0}^*$, and \(\phi_{s_{\mathrm{est},i}}\) is a complex zero-mean Gaussian noise\cite{ref14}. After the traversals of $i$ and $\eta$ are completed, the estimations of modulated symbols $\hat{\textbf{D}}=\{\hat{d}_1,...,\hat{d}_{\kappa},...,\hat{d}_{N_{\mathrm{tsu}}}\}$ are generated. Finally, modulated symbols $\hat{\textbf{D}}$ are converted to information bits $\{\hat{\tau}_1,...,\hat{\tau}_{\kappa},...,\hat{\tau}_{N_{\mathrm{tsu}}}\}$. In general, the demodulation process for the proposed SFI-LoRa scheme is listed in Algorithm \ref{demodulate}.

{\em Remark 1:} Since the structure of data frame in the proposed SFI-LoRa scheme is not modified, the MAC protocol  \cite{ref76} and the ALOHA protocol \cite{ref77} used in existing LoRa schemes can also be used in the SFI-LoRa scheme. However, a lower duty cycle can be set in the proposed SFI-LoRa scheme compared to the conventional LoRa scheme because the SFI-LoRa scheme has a shorter air time when transmitting the same number of information bits.

{\em Remark 2:} To implement the proposed SFI-LoRa scheme, slight hardware changes are required at both the transmitter and receiver.

\section{Analysis of Performance and Efficiency}
Generally speaking, the averaged SER of the proposed SFI-LoRa scheme is attributed to the erroneous detection of index symbol and modulated symbols. In the following, we will first derive the SER expressions of the index symbol and modulated symbols and then formulate the averaged SER of the proposed SFI-LoRa scheme. It should be noted that we assume the LoRa signals with different SFs are perfect orthogonality in the theoretical analysis for simplicity, because the amplitude of the sidelobes of LoRa signals is trivial and can be ignored \cite{ref103,ref104,ref105}.
\subsection{SER Performance}
%

\subsubsection{SER Analysis of Index Symbol}
Here, we define the \(N_\mathrm{av}-M\) SFs that are not selected as $\textbf{S}_{\mathrm{ns}}$ (i.e., $\textbf{S}_{\mathrm{av}}=\textbf{S}_{\mathrm{ns}}\cup \textbf{S}_{\delta}$), where \(\textbf{S}_{\mathrm{ns}}=\{s_{\mathrm{ns},1},...,s_{\mathrm{ns},j},...,s_{\mathrm{ns},N_\mathrm{av}-M}\}\).
In the proposed SFI-LoRa scheme, the SER of index symbol \(P_{\mathrm{ise}}\) is derived by the pairwise error probabilities of between the estimated SF combination \(\textbf{S}_\mathrm{est}\) and $\textbf{S}_{\mathrm{ns}}$. The detection of \(\textbf{S}_\mathrm{est}\) depends on the $N_{\mathrm{av}}$ maximum amplitudes \textbf{c} that exhibits the correlation with the received signal. It should be emphasized that the interception of the received signal during the detection of the index symbol contains multiple incomplete LoRa signals with different SFs. Since the amplitude of the side lobes is much lower than that of the main lobe, the influence of the side lobes are neglected in the analysis. Moreover, according to (\ref{f22}), all elements in \textbf{c} have two possible values. In the first case, $c_\nu=\sqrt{\frac{h E_{\mathrm{s}}}{M\cdot 2^{i-1}\cdot 2^{s_{\mathrm{est},i}}}}+\phi_{s_{\mathrm{av},\nu}}$ when $\lambda=d_{t}$ and $f\in \textbf{S}_{\delta}$. In the second case, $c_\nu=\phi_{s_{\mathrm{av},\nu}}$ when $\lambda\neq d_{t}, f\in \textbf{S}_{\delta}$ or $f\notin \textbf{S}_{\delta}$. Therefore, the pairwise error probability $P_{\mathrm{pe}|i,j,h}$ of \(s_{\delta,i}\) and $s_{\mathrm{ns},j}$ has two components, which can be expressed as
\begin{equation}\label{f23}
\begin{aligned}
    P_{\mathrm{pe}|i,j,h}=&\mathrm{Pr}[|R_{s_{\mathrm{ns},j},1}|>|R_{s_{\mathrm{est},i},1}|]\\
=&(1-P_{\mathrm{sub},i})P_{\mathrm{ind,\uppercase\expandafter{\romannumeral1}}}+P_{\mathrm{sub},i}P_{\mathrm{ind,\uppercase\expandafter{\romannumeral2}}}\\
    =&(1-P_{\mathrm{sub},i})\mathrm{Pr}[\mathrm{max}(|\phi_{s_{\mathrm{ns},j}}|)>\beta_{s_{\mathrm{est},i}}]\\
    &+P_{\mathrm{sub},i}\mathrm{Pr}[\mathrm{max}(|\phi_{s_{\mathrm{ns},j}}|)>|\phi_{s_{\mathrm{est},i}}|],
\end{aligned}
\end{equation}
where \(\beta_{s_{\mathrm{est},i}}=\bigg|\sqrt{\frac{h E_{\mathrm{s}}}{M\cdot 2^{i-1}\cdot 2^{s_{\mathrm{est},i}}}}+\phi_{s_{\mathrm{est},i}}\bigg|\) is the magnitude of the complex noise envelope, $P_{\mathrm{sub},i}$ is the SER of modulated symbol in $i$-th block and \(\mathrm{max}(\cdot)\) returns maximum value of the parameters. It should be noted that $\beta_{s_{\mathrm{est},i}}$ follows a Rice distribution with the shape parameter of $\kappa=E_\mathrm{s}/2\sigma^2=E_\mathrm{s}/N_0$.
The maximum value of the \(2^{f}\) independent and identically distributed Rayleigh random variables $|\phi_{f}|$ is indicated by \(\rho_{f}=\mathrm{max}(|\phi_{f}|)\).
The cumulative distribution function (CDF) $\Lambda_{\rho_{f}}(\rho_f)$ and probability density function (PDF) $\lambda_{\rho_{f}}(\rho_f)$ of \(\rho_f\) can be respectively given as \cite{ref54}
\begin{equation}\label{f24}\Lambda_{\rho_{f}}(\rho_f)=\left[1-\exp\left(-\frac{\rho_f^{2}}{2\sigma^{2}}\right)\right]^{2^{f}},\end{equation}
\begin{equation}\label{f30}
\begin{aligned}
\lambda_{\rho_{f}}(\rho_f)&\!=\!{2^{f}}\left[1\!-\!\exp\left(\!-\!\frac{\rho_f^{2}}{2\sigma^{2}}\right)\right]^{2^{f}-1}\!\cdot\!\frac{\rho_f}{\sigma^{2}}\exp\left(-\frac{\rho_f^{2}}{2\sigma^{2}}\right).
\end{aligned}
\end{equation}
Base on (\ref{f24}), the SER $P_{\mathrm{ind,\uppercase\expandafter{\romannumeral1}}}$ of the first case can be further formulated as \cite{ref14}
\begin{equation}\label{f25}
\begin{aligned}
P_{\mathrm{ind,\uppercase\expandafter{\romannumeral1}}}=&\mathrm{Pr}[\mathrm{max}(|\phi_{s_{\mathrm{ns},j}}|)>\beta_{s_{\mathrm{est},i}}]\\=&\int_{0}^{\infty}\left[1-\left[1-\exp\left(-\frac{\beta_{s_{\mathrm{est},i}}^{2}}{2\sigma^{2}}\right)\right]^{2^{s_{\mathrm{ns},j}}}\right]\\&\times  f_{\beta_{s_{\mathrm{est},i}}}(\beta_{s_{\mathrm{est},i}})d\beta_{s_{\mathrm{est},i}},
\end{aligned}
\end{equation}
where the PDF \(f_{\beta_{s_{\mathrm{est},i}}}(\beta_{s_{\mathrm{est},i}})\) for the Rice distributed \(\beta_{s_{\mathrm{est},i}}\) is represented by
\begin{equation}\label{f26}
{f_{{\beta _{s_{\mathrm{est},i}} }}}\left ({{{\beta _{s_{\mathrm{est},i}} }} }\right) = \frac {{2{\beta _{s_{\mathrm{est},i}} }}}{{{N_{0}}}}{I_{0}}\left ({{\frac {{2\xi_i{\beta _{s_{\mathrm{est},i}} }}}{{{N_{0}}}}} }\right){e^{ - \frac {{\beta _{s_{\mathrm{est},i}}^{2} + \xi_i^{2}}}{{{N_{0}}}}}}.
\end{equation}
In (\ref{f26}), $I_{\varphi}(z)\!=\!\frac{1}{\pi} \int_0^\pi e^{z \cos \theta} \cos(\varphi\theta) d\theta$ denotes the $\varphi$th-order modified Bessel function of the first kind \cite{ref53}, and $\xi_i=\sqrt{\frac{h E_{\mathrm{s}}}{M\cdot 2^{i-1}\cdot 2^{s_{\mathrm{est},i}}}}$. Substituting (\ref{f26}) into (\ref{f25}) yields
\begin{equation}\begin{aligned}\label{f27}
P_{\mathrm{ind,\uppercase\expandafter{\romannumeral1}}} =& \int_{0}^{\infty}\left[1-\left[1-\exp\left(-\frac{\beta_{s_{\mathrm{est},i}}^{2}}{N_{0}}\right)\right]^{2^{s_{\mathrm{ns},j}}}\right] \\
 &\times \frac {{2{\beta _{s_{\mathrm{est},i}} }}}{{{N_{0}}}}{I_{0}}\left ({{\frac {{2\xi_i{\beta _{s_{\mathrm{est},i}} }}}{{{N_{0}}}}} }\right){e^{ - \frac {{\beta _{s_{\mathrm{est},i}}^{2} + \xi_i^{2}}}{{{N_{0}}}}}}d\beta_{s_{\mathrm{est},i}} \\
  =&\sum_{k=1}^{2^{s_{\mathrm{ns},j}}}(-1)^{k+1}\binom{2^{s_{\mathrm{ns},j}}}{k}e^{-\frac{k\xi_i^2}{(k+1)N_0}} \int_{0}^{\infty}\frac{2\beta_{s_{\mathrm{est},i}}}{N_{0}}\\& \times I_{0}\left(\frac{2\xi_i\beta_{s_{\mathrm{est},i}}}{N_{0}}\right)e^{-\frac{(k+1)\beta_{s_{\mathrm{est},i}}^{2}+\frac{\xi_i^{2}}{k+1}}{N_{0}}}d\beta_{s_{\mathrm{est},i}}.
\end{aligned}\end{equation}
Let $\tilde{\xi}_i=\frac{\xi_i}{\sqrt{k+1}}$ and $\tilde{\beta}_{s_{\mathrm{est},i}}\!=\!\sqrt{k+1}\beta_{s_{\mathrm{est},i}}$, (\ref{f27}) can be rewritten as
\begin{equation}\label{f28}
\begin{aligned}
P_{\mathrm{ind,\uppercase\expandafter{\romannumeral1}}}  =&\sum_{k=1}^{2^{s_{\mathrm{ns},j}}}(-1)^{k+1}\binom{2^{s_{\mathrm{ns},j}}}{k}e^{-\frac{k\xi_i^{2}}{(k+1)N_{0}}}   \frac{1}{k+1}\\
&\times\int_{0}^{\infty}\frac{2\tilde{\beta}_{s_{\mathrm{est},i}}}{N_{0}}I_{0}\biggl(\frac{2\tilde{\xi}_i\tilde{\beta}_{s_{\mathrm{est},i}}}{N_{0}}\biggr)e^{-\frac{\tilde{\xi}_i^{2} + \tilde{\beta}^{2}_{s_{\mathrm{est},i}}}{N_{0}}}d\tilde{\beta}_{s_{\mathrm{est},i}} \\
 =&\sum_{k=1}^{s_{\mathrm{ns},j}}\frac{(-1)^{k+1}}{k+1}\binom{2^{s_{\mathrm{ns},j}}}{k}e^{-\frac{k}{(k+1)M\cdot 2^{i-1}\cdot2^{s_{\mathrm{est},i}}}\cdot\frac{h E_{s}}{N_{0}}}.
\end{aligned}\end{equation}
Similar to the derivation of (\ref{f25}), the SER $P_{\mathrm{ind,\uppercase\expandafter{\romannumeral2}}}$ of the second case can be further caculated by
\begin{equation}\label{f29}
\begin{aligned}
P_{\mathrm{ind,\uppercase\expandafter{\romannumeral2}}}=&\mathrm{Pr}\left[\mathrm{max}(|\phi_{s_{\mathrm{ns},j}}|)>|\phi_{s_{\mathrm{est},i}}|\right]\\
=&\int_{0}^{\infty}\left[1-\left[1-\exp\left(-\frac{\rho^{2}}{2\sigma^{2}}\right)\right]^{2^{s_{\mathrm{ns},j}}}\right]\\&\times  \lambda_{\rho_{s_{\mathrm{est},i}}}(\rho_{s_{\mathrm{est},i}})d\rho\\
=&\int_{0}^{\infty}\left[1-\left[1-\exp\left(-\frac{\rho^{2}}{2\sigma^{2}}\right)\right]^{2^{s_{\mathrm{ns},j}}}\right]\\&\times
{2^{s_{\mathrm{est},i}}}\left[1-\exp\left(-\frac{\rho^{2}}{2\sigma^{2}}\right)\right]^{2^{s_{\mathrm{est},i}}-1}\\&\times\frac{\rho}{\sigma^{2}}\exp\left(-\frac{\rho^{2}}{2\sigma^{2}}\right) d\rho.
\end{aligned}
\end{equation}
Let $\zeta=1-\exp(-\frac{\rho^2}{2\sigma^2})$ and $d\zeta=\frac{\rho}{\sigma^2}\exp(-\frac{\rho^2}{2\sigma^2})d\rho$, (\ref{f29}) can be written as
\begin{equation}
\begin{aligned}
P_{\mathrm{ind,\uppercase\expandafter{\romannumeral2}}}&=2^{s_{\mathrm{est},i}}\int_{0}^{1}\left(1-\zeta^{2^{s_{\mathrm{ns},j}}}\right)\times  \zeta^{2^{s_{\mathrm{est},i}}-1}d\zeta\\&
=2^{s_{\mathrm{est},i}}\left[\int_{0}^{1} \zeta^{2^{s_{\mathrm{est},i}}-1}d\zeta-\int_{0}^{1} \zeta^{2^{s_{\mathrm{ns},j}}+2^{s_{\mathrm{est},i}}-1}d\zeta\right] \\&
=\zeta^{2^{s_{\mathrm{est},i}}}\Big|^1_0-\frac{2^{s_{\mathrm{est},i}}}{2^{s_{\mathrm{ns},j}}+2^{s_{\mathrm{est},i}}}\zeta^{2^{s_{\mathrm{ns},j}}+2^{s_{\mathrm{est},i}}}\Big|^1_0\\&
=\frac{2^{s_{\mathrm{ns},j}}}{2^{s_{\mathrm{ns},j}}+2^{s_{\mathrm{est},i}}}.
\end{aligned}
\end{equation}

This paper considers the frequency-flat Rayleigh fading channel that has previously been widely employed in the context of LoRa communication scheme studies \cite{ref11,ref94,ref95}. Since $\sqrt{h}$ is a Rayleigh random variable, $h$ follows a Chi-squared distribution with two degrees of freedom. Hence, the PDF of $h$ can be expressed as $f_h(h)=e^{-h}$. Moreover, it is necessary to obtain an estimation of the SF combination \(\textbf{S}_\mathrm{est}=\{s_{\mathrm{est},1},...,s_{\mathrm{est},i},...,s_{\mathrm{est},M}\}\) to recover index bits. The number of elements in $\textbf{S}_{\mathrm{est}}$ and $\textbf{S}_{\mathrm{ns}}$ are $M$ and $N_{\mathrm{av}}-M$, respectively. Particularly, each element in the $\textbf{S}_{\mathrm{est}}$ is compared to each element in $\textbf{S}_{\mathrm{ns}}$, hence $M(N_{\mathrm{av}}-M)$ cases need to be considered. Eventually, the SER of the index symbol can be derived as
\begin{equation}
P_{\mathrm{ise}}=1-\prod_{\tau=1}^{M}\prod_{\gamma=1}^{N_{\mathrm{av}}-M}(1-P_{\mathrm{pe}|\tau,\gamma,h}).
\end{equation}


\subsubsection{SER Analysis of Modulated Symbols}

In the proposed SFI-LoRa scheme, because the detection of modulated symbols in different sub-blocks are independent of one another, the error probability of the modulated symbol carried in each sub-block are identical. The SER of the $i$-th block is expressed as \cite{ref45}
\begin{equation}\label{f32}
\begin{aligned}
P_{\mathrm{sub},i}=&\int_{0}^{\infty}\left[1-\left[1-\exp\left(-\frac{\beta_{s_{\mathrm{est},i}}^{2}}{2\sigma^{2}}\right)\right]^{2^{s_{\mathrm{est},i}-1}}\right]
\\&\times  f_{\beta_{s_{\mathrm{est},i}}}(\beta_{s_{\mathrm{est},i}})d\beta_{s_{\mathrm{est},i}}
\\=&\sum_{k=1}^{2^{s_{\mathrm{est},i}}}(-1)^{k+1}\binom{2^{s_{\mathrm{est},i}}}{k}e^{-\frac{k\xi_i^2}{(k+1)N_0}} \int_{0}^{\infty}\frac{2\beta_{s_{\mathrm{est},i}}}{N_{0}}\\& \times I_{0}\left(\frac{2\xi_i\beta_{s_{\mathrm{est},i}}}{N_{0}}\right)e^{-\frac{(k+1)\beta_{s_{\mathrm{est},i}}^{2}+\frac{\xi_i^{2}}{k+1}}{N_{0}}}d\beta_{s_{\mathrm{est},i}}
\\=&\sum_{k=1}^{2^{s_{\mathrm{est},i}}-1}\frac{\left(-1\right)^{k+1}}{k+1}\binom{2^{s_{\mathrm{est},i}}-1}{k}\\&\times\exp\left[-\frac{k}{k+1}\cdot\frac{hE_{S}}{2^{s_{\mathrm{est},i}}\!\cdot \! 2^{i-1}\!\cdot\!MN_{0}}\right].
\end{aligned}
\end{equation}
Based on the number of sub-blocks in the $i$-th block $N_{\mathrm{sub},i}$ and (\ref{f32}), the modulated SER of the transmitted signal is obtained as
\begin{equation}P_{\mathrm{ms}}=1-\prod_{i=1}^{M}P_{\mathrm{sub},i}^{N_{\mathrm{sub},i}}.\end{equation}

Finally, the SER for the proposed SFI-LoRa scheme can be written as
\begin{equation}
P_\mathrm{s}=\int_{0}^{\infty}\left[1-(1-P_{\mathrm{ise}})(1-P_{\mathrm{ms}})\right]f_{h}(h)dh.
\end{equation}

\begin{table}[t]
\centering
\renewcommand\arraystretch{1.85}
\caption{\textsc{Data Rate Comparison of the state-of-the-art LoRa Schemes}}
\label{tab3}
\begin{tabular}{|m{2cm}<{\centering}|m{5cm}<{\centering}|}

\hline
Scheme & Data Rate    \\ \hline

LoRa  \cite{ref4} &  $\displaystyle\dfrac{f}{2^f\cdot T_{\mathrm{chip}}}$    \\ \hline

ICS-LoRa \cite{ref23} & $\displaystyle\dfrac{f+1}{2^f\cdot T_{\mathrm{chip}}}$    \\ \hline

SSK-LoRa \cite{ref24}& $\displaystyle\dfrac{f+1}{2^f\cdot T_{\mathrm{chip}}}$    \\ \hline

PSK-LoRa \cite{ref9}& $\displaystyle\dfrac{f+N_{\mathrm{p}}}{2^f\cdot T_{\mathrm{chip}}}$  \\ \hline



MuLoRa \cite{ref10}& $\displaystyle\dfrac{2^{SF_\mathrm{a}}(f-SF_\mathrm{a})}{2^f\cdot T_{\mathrm{chip}}}$    \\ \hline

DCDSK \cite{ref55}& $\displaystyle\dfrac{2f}{2^f\cdot T_{\mathrm{chip}}}$    \\ \hline
%

\end{tabular}
\end{table}

\subsection{Data Rate}
In this subsection, we analyze the data rate of the proposed SFI-LoRa scheme. The data transmission rates for the conventional LoRa \cite{ref4}, ICS-LoRa \cite{ref23}, SSK-LoRa \cite{ref24}, PSK-LoRa \cite{ref9}, FBI-LoRa \cite{ref11}, MuLoRa \cite{ref10}, HM-LoRa \cite{ref93}, and DCDSK \cite{ref55} schemes are summarized in Table \ref{tab3}. It should be noted that $N_{\mathrm{p}}$ represents the PSK bits in the PSK-LoRa, while $2^{SF_{\mathrm{a}}}$ LoRa symbols is transmitted during a symbol duration in MuLoRa.

The data rate is typically determined by the ratio of the number of information bits conveyed each symbol to the symbol duration \cite{ref56,ref68,ref58}. Based on the analysis above, the number of information bits in the proposed SFI-LoRa scheme is expressed as
\begin{equation}
N_{\mathrm{tot}}=\left\lfloor\log_2\begin{pmatrix}N_{\mathrm{av}}\\M\end{pmatrix}\right\rfloor+\sum_{i=1}^{M}s_{\delta,i}\cdot N_{\mathrm{sub},i}.
\end{equation}
Furthermore, the symbol duration can be calculated by $2^{s_{\delta,1}}\cdot T_{\mathrm{chip}}$, hence the data rate of the proposed scheme is computed as
\begin{equation}
\Theta=\frac{\left\lfloor\log_2\begin{pmatrix}N_{\mathrm{av}}\\M\end{pmatrix}\right\rfloor+\sum_{i=1}^{M}s_{\delta,i}\cdot N_{\mathrm{sub},i}}{\sum_{i=1}^{M}\sum^{N_{\mathrm{com}}}_{\delta=1}2^{s_{\delta,i}+i-1}/(MN_{\mathrm{com}})\cdot T_{\mathrm{chip}}}.
\end{equation}
To facilitate the comparison, we assume that the parameter settings for the proposed SFI-LoRa, PSK-LoRa, MuLoRa, DCDSK, and HM-LoRa  are $M=2$, $N_{\mathrm{p}}=2$, $SF_{\mathrm{a}}=1$, $f=9$, and $\textbf{S}_{\mathrm{Hm}}=\{s_{\mathrm{Hm},1},s_{\mathrm{Hm},2}\}=\{9,8\}$, respectively.\footnote{Since the minimum number of superimposed signals within a symbol duration for DCDSK and MuLoRa are $2$, we set the number of selected SFs $M = 2$. Moreover, the average symbol duration of the proposed SFI-LoRa scheme is $2^{9.187}T_\mathrm{chip}$ when $M = 2$, which is similar to the competitors at $f = 9$.} For the similar average symbol duration, the data rate of the proposed SFI-LoRa is $\frac{28}{\sum_{i=1}^{M}\sum^{N_{\mathrm{com}}}_{\delta=1}2^{s_{\delta,i}+i-1}/(MN_{\mathrm{com}})\cdot T_{\mathrm{chip}}}$. Based on the above parameter setting, the proposed SFI-LoRa, conventional LoRa, ICS/SSK LoRa, PSK-LoRa, MuLoRa, DCDSK, and HM-LoRa can achieve the data rates of 6.003kbps, 2.197kbps, 2.441kbps, 2.685kbps, 3.906kbps, 4.394kbps, and 5.164kbps, respectively. The data rate of the proposed SFI-LoRa scheme is 173\%, 145\%, 123\%, 53\%, 36\%, 14\% higher than that of the conventional LoRa, ICS/SSK-LoRa, PSK-LoRa, MuLoRa, DCDSK, and HM-LoRa schemes, respectively.  As mentioned above, the proposed SFI-LoRa scheme offers a substantially improved data rate relative to the existing competitors. 

\subsection{Transmission Throughput}
To demonstrate the superiority of the proposed SFI-LoRa scheme, it is essential to compare its throughput with other LoRa schemes for a comprehensive evaluation. It should be noted that the proposed scheme must undergo a zero padding before transmission to maintain a consistent symbol duration. Therefore, we will conduct the theoretical analyses and simulations based on this processing method.

In the context of wireless communication systems, the throughput is defined as the quantity of information bits that can be accurately detected by the receiving end. This concept is expressed as follows \cite{ref11}
\begin{equation}
T_\mathrm{r}=\frac{(N_{\mathrm{p}}N_b\left(1-P_{\mathrm{per}}\right))}{T_\mathrm{p}}.
\end{equation}
Here, the number of symbols in each packet is denoted as \(N_{\mathrm{p}}\), and \(N_b\) represents the number of information bits carried by each symbol. The transmission period of each packet is given by \(T_\mathrm{p}=N_{\mathrm{p}}\cdot2^{f}\cdot T_{\mathrm{chip}}\), and \(P_{\mathrm{per}}\) corresponds to the packet error rate, which is given by
\begin{equation}P_{\mathrm{per}}=1-(1-P_\mathrm{s})^{N_{\mathrm{p}}}.\end{equation}

\subsection{Computational and Space Complexities}

In this paper, the receiver complexity of the proposed SFI-LoRa scheme consists of computational complexity and space complexity. The computational complexity of the SFI-LoRa scheme mainly consists of two components: i) the complexity of index search and ii) the complexity of LoRa demodulation. During the index search, the receiver must identify the SFs combination. The index search includes dechirp operation, DFT operation, absolute value operation, and peak detection. The computational complexities of dechirp operation, DFT operation, absolute value operation, and peak detection are $\mathcal{O}(\eta_{s_{\mathrm{opt},\mu}})$, $\mathcal{O}(\eta^2_{s_{\mathrm{opt},\mu}})$,$\mathcal{O}(\eta_{s_{\mathrm{opt},\mu}})$, and $\mathcal{O}(\eta_{s_{\mathrm{opt},\mu}})$, respectively, where $\eta_{i}=2^{i}$. Therefore, the computational complexity of index search can be expressed by $\mathcal{O}{\left(\sum_{i=1}^{N_{\mathrm{opt}}}\eta_{s_{\mathrm{opt},i}}^2+3\eta_{s_{\mathrm{opt},\mu}}\right)}$.  During the LoRa demodulation, although the SFI-LoRa scheme uses the same mapping method as the conventional LoRa scheme to convey modulated bits, the transmitted signal of the SFI-LoRa scheme contains $N_\mathrm{tsu}$ modulated symbols and $M$ different SFs. Therefore, based on the computational complexity of demodulation in the conventional LoRa scheme, i.e., $\mathcal{O}(\eta_f^2+3\eta_f)$ \cite{ref31}, the computational complexity of LoRa demodulation in the SFI-LoRa scheme can be represented by $\mathcal{O}\left(\sum_{\mu=0}^{M-1}[2^\mu(\eta_{s_{\mathrm{est},\mu}}^2+3\eta_{s_{\mathrm{est},\mu}})]\right)$. Similarly, the computational complexities of the MuLoRa, DCDSK, HM-LoRa, FBI-LoRa schemes are given by $\mathcal{O}(\eta_f^2+3\eta_f)$, $\mathcal{O}(2\eta_f^2+6\eta_f)$,  $\mathcal{O}(\sum_{i=1}^{K}2^{i-1}[\eta_{s_{\mathrm{Hm},i}}^2+\eta_{s_{\mathrm{Hm},i}}])$, and $\mathcal{O}(\eta_f^2+3\eta_f+f_\mathrm{num}g_\mathrm{num}N_\mathrm{ac})$, respectively, where $f_\mathrm{num}$ denotes the number of the SFB in each group, and $g_\mathrm{num}$ denotes the number of indices groups in the FBI-LoRa scheme. Obviously, the computational complexity of the proposed SFI-LoRa scheme is the same order as those of existing LoRa schemes, i.e,  $\mathcal{O}{(\eta_f^2)}$.
On the other hand, the space complexity can be treated as the memory requirement \cite{ref83}. At the receiver of our design, the memory requirements of the index search and LoRa demodulation are $\mathcal{O}\left(\sum_{i=0}^{N_\mathrm{opt}-1}3\eta_{s_{\mathrm{opt},i}}\right)$ and $\mathcal{O}\left(\sum_{i=0}^{M-1}3\eta_{s_{\mathrm{est},i}+i}\right)$, respectively. Therefore, the space complexity of the SFI-LoRa scheme can be expressed as $\mathcal{O}\left[\left(\sum_{i=0}^{N_\mathrm{opt}-1}3\eta_{s_{\mathrm{opt},i}}\right)+\left(\sum_{i=0}^{M-1}3\eta_{s_{\mathrm{est},i}+i}\right)\right]$. Likewise, the space complexities of the conventional LoRa, MuLoRa, DCDSK, HM-LoRa and FBI-LoRa schemes are calculated as $\mathcal{O}{(3\eta_f)}$, $\mathcal{O}{(3\eta_f)}$, $\mathcal{O}{(5\eta_f)}$, $\mathcal{O}({\sum_{i=1}^{K}2^{i-1}3\eta_{s_{\mathrm{Hm},i}}})$, $\mathcal{O}{(3f_\mathrm{num}g_\mathrm{num}\eta_f)}$. Thereby, the space complexity of the SFI-LoRa scheme is also the same order as those of the conventional LoRa, MuLoRa, DCDSK, HM-LoRa, and FBI-LoRa schemes i.e., $\mathcal{O}{(\eta_f)}$.


\begin{table}[t]
\centering
\vspace{0cm}
\footnotesize{\centering ~~~~~TABLE~V. Energy efficiencies of the proposed SFI-LoRa, MuLoRa, DCDSK, HM-LoRa, and FBI-LoRa schemes.
    }\vspace{+2mm}
\label{energy}
\renewcommand{\arraystretch}{1.3}
\begin{tabular}{|m{1.5cm}<{\centering}|m{3.5cm}<{\centering}|}

\hline
Scheme &  Energy efficiency  \\ \hline

LoRa  &   $\frac{f}{2^f}$  \\ \hline

MuLoRa  &   $\frac{2^{SF_\mathrm{a}}(f-SF_\mathrm{a})}{2^f}$  \\ \hline

DCDSK  &   $\frac{2\cdot f}{2^f}$  \\ \hline

HM-LoRa &    $\frac{\sum_{k=1}^{K}2^{s_{\mathrm{Hm},1}-s_{\mathrm{Hm},k}}s_{\mathrm{Hm},k}}{2^{s_\mathrm{{Hm},1}}}$   \\ \hline

FBI-LoRa &     $ \frac{g_\mathrm{num}\left\lfloor \log_2\left(\begin{smallmatrix}2^f/g_\mathrm{num}\\f_\mathrm{num}\end{smallmatrix}\right)\right\rfloor}{2^f} $  \\ \hline

SFI-LoRa  &  $\sum_{i=1}^{M}\frac{s_{\delta,i}N_{\mathrm{sub},i}}{2^{s_{\delta,1}}}$ \\ \hline

\end{tabular}
\end{table}

\subsection{Energy Efficiency}
The energy efficiency of a LoRa scheme can be defined as the ratio of the information-bit number carried by each modulated symbol to the average symbol energy \cite{ref98,ref99}. In this paper, the energy of a single chip in the LoRa signal is assumed as $E_\mathrm{c}=1$. Hence, the energy efficiency of the conventional LoRa, MuLoRa, DCDSK, HM-LoRa, and FBI-LoRa schemes can be calculated in Table.~V. More specifically, based on the parameter setting in Section IV-B, the energy efficiency of the proposed SFI-LoRa, conventional LoRa, MuLoRa, DCDSK, HM-LoRa, and FBI-LoRa schemes are 0.0481, 0.017, 0.031, 0.035, 0.0488, and 0.054, respectively. Obviously, the energy efficiency of the proposed SFI-LoRa scheme is higher than that of the conventional LoRa, MuLoRa, and DCDSK schemes, but is slightly lower than that of the HM-LoRa and FBI-LoRa schemes. In consequence, the proposed SFI-LoRa scheme can achieve the better trade-off between the BER performance and throughput performance with a desirable energy efficiency. As such, our design can well satisfy high-data-rate wireless communication applications which are relatively less sensitive to energy consumption \cite{ref100,ref101}.

\begin{figure}[t]
\centering
\includegraphics[width=3.3 in]{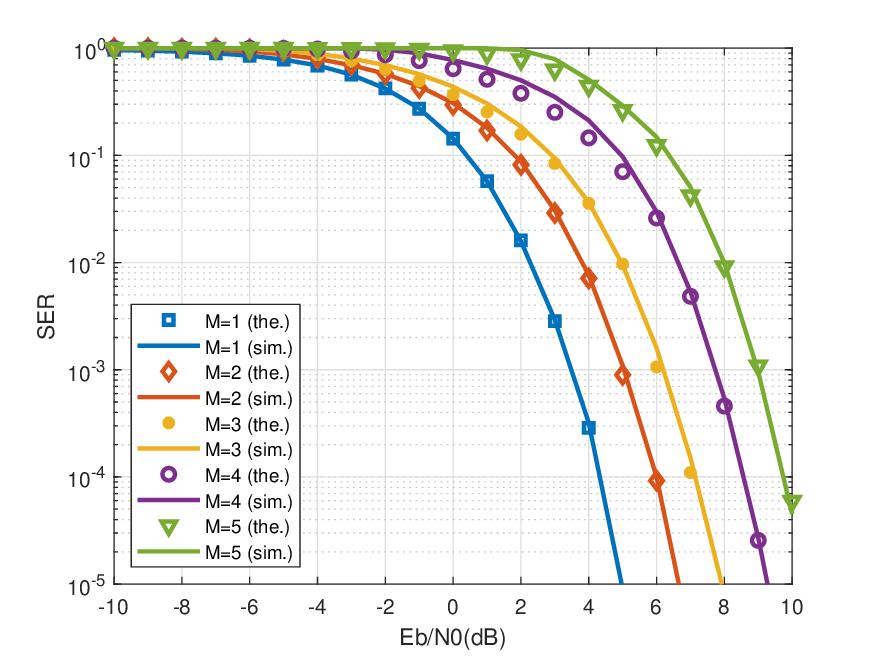}
\caption{ Theoretical and simulated SER results of the proposed SFI-LoRa over an AWGN channel.}
\label{fig3}
\end{figure}
\begin{figure}[t]
\centering
\includegraphics[width=3.3 in]{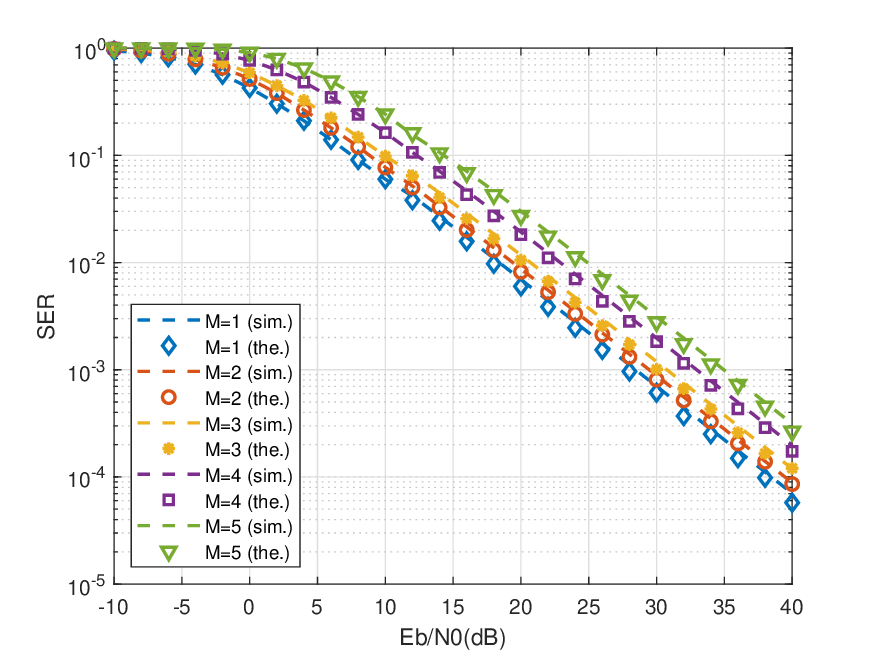}
\caption{ Theoretical and simulated SER results of the proposed SFI-LoRa over a Rayleigh fading channel.}
\label{fig4}
\end{figure}

\section{Simulation Results and Discussion}
In this section, we utilize numerical simulation for estimating the SER and throughput performance of the proposed SFI-LoRa scheme over (single-path) AWGN and Rayleigh fading, and multipath flat-fading AWGN channels.
Unless otherwise mentioned, the parameter setting for the proposed scheme and other benchmarks in this section is the same as in Section~IV-B, and a two-path flat-fading channel (i.e., C2-channel in \cite{ref110}) is assumed in the corresponding simulations.

\begin{figure*}[t]
\centering
\subfigure[\hspace{-0.5cm}]{\label{fig:Capacity-5slots}
\includegraphics[width=2.3in,height=2.0in]{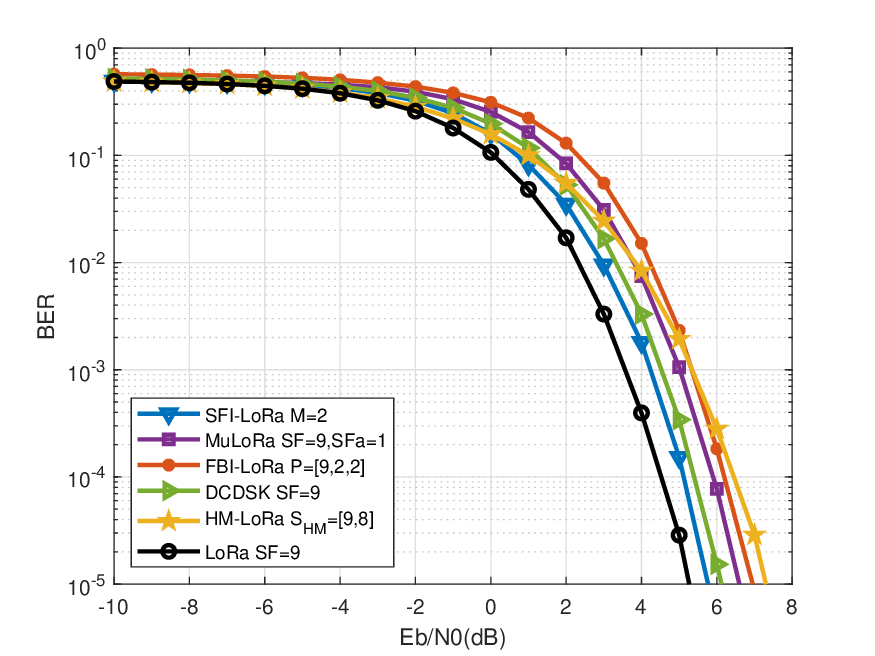}}
\subfigure[\hspace{-0.5cm}]{\label{fig:Capacity-6slots}
\includegraphics[width=2.3in,height=2.0in]{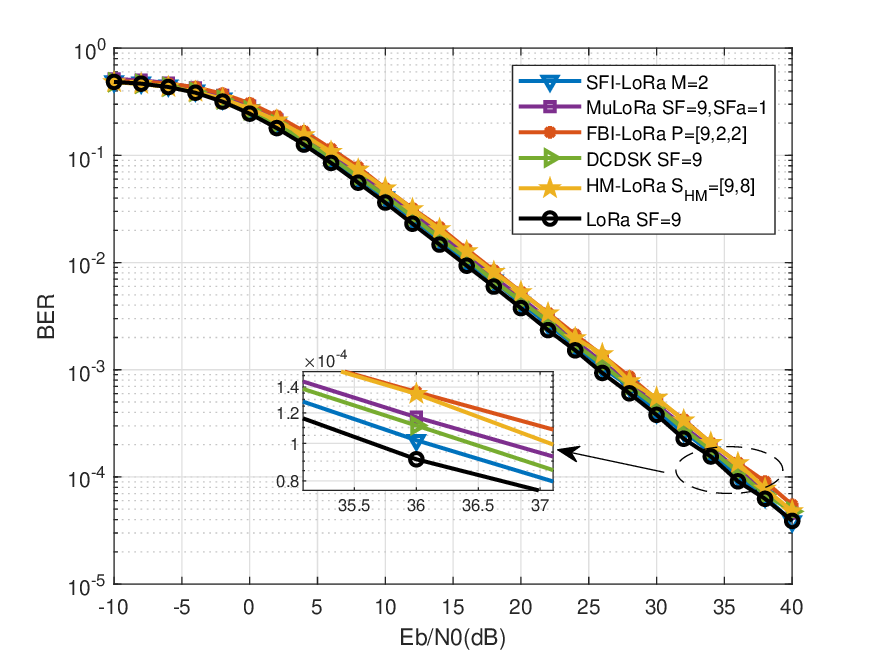}}
\subfigure[\hspace{-0.5cm}]{\label{fig:Capacity-6slots}
\includegraphics[width=2.3in,height=2.0in]{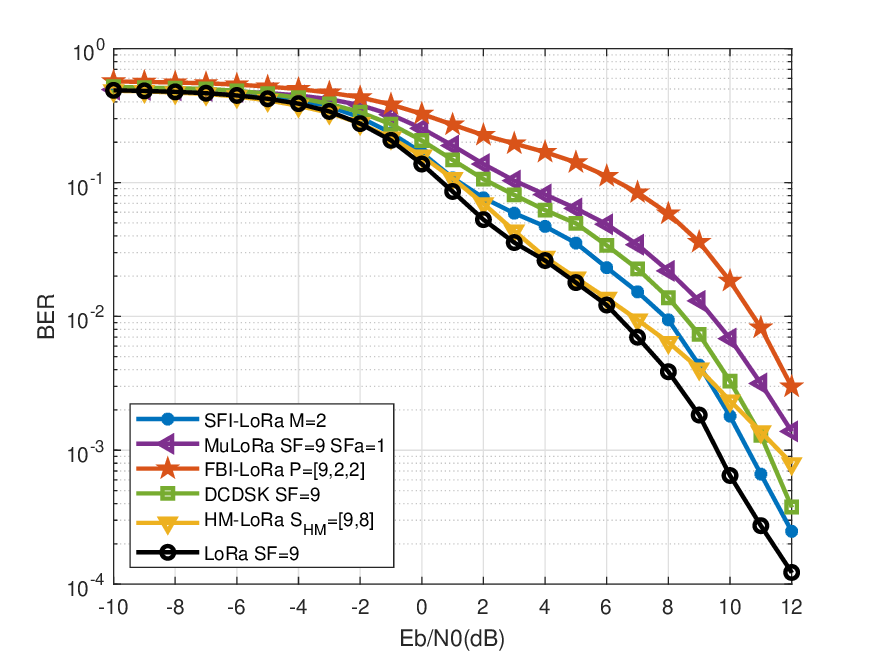}}
\caption{The BER performance of the proposed SFI-LoRa scheme compared with MuLoRa, FBI-LoRa, HM-LoRa and DCDSK schemes over the (a) AWGN, (b) single-path Rayleigh fading, and (c) two-paths flat-fading AWGN channels ($h_\mathrm{m}=0.7$ and $n_i=1$).}
\label{fig:BER-compare}
\end{figure*}

\vspace{-2mm}
\begin{figure*}[t]
\centering
\subfigure[\hspace{-0.5cm}]{\label{fig:Capacity-5slots}
\includegraphics[width=2.3in,height=2.0in]{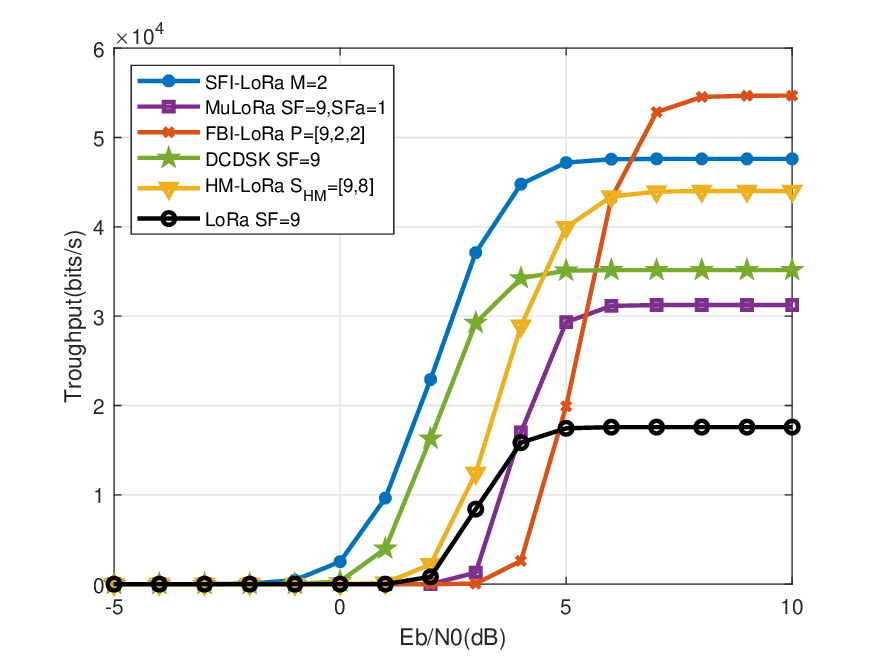}}
\subfigure[\hspace{-0.5cm}]{\label{fig:Capacity-6slots}
\includegraphics[width=2.3in,height=2.0in]{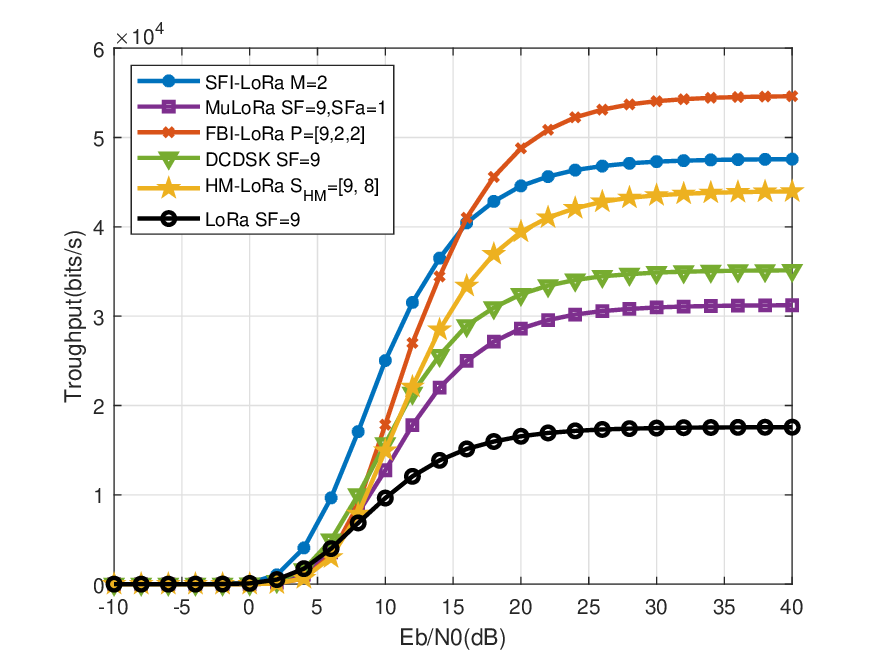}}
\subfigure[\hspace{-0.5cm}]{\label{fig:Capacity-6slots}
\includegraphics[width=2.3in,height=2.0in]{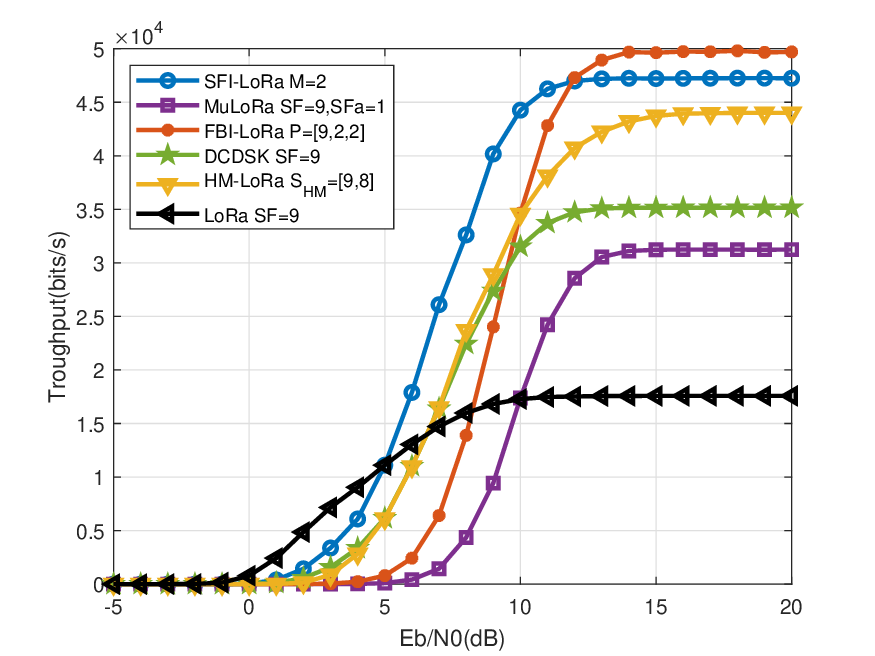}}\vspace{-2mm}
\caption{The BER performance of the proposed SFI-LoRa scheme compared with DCDSK, MuLoRa, HM-LoRa, and FBI-LoRa schemes over the (a) AWGN, (b) single-path Rayleigh fading, and (c) two-paths flat-fading AWGN channels ($h_\mathrm{m}=0.7$ and $n_i=1$).}
\label{fig:throughput-compare}
\end{figure*}

\subsection{Simulated and Theoretical Performance}

Figs.~\ref{fig3} and \ref{fig4} present the theoretical and simulation results of the proposed SFI-LoRa scheme. The term ``sim." denotes the simulation result with imperfect SF orthogonality,\footnote{In the scenario of imperfect SF orthogonality, the LoRa signal with a given SF suffers from the interference from other LoRa signals with different SFs.} whereas ``the." indicates the theoretical analysis with perfect SF orthogonality assumption. As observed, a high consistency can be observed between the theoretical SER curves and the simulated ones across various parameter settings. In particular, the theoretical SER results of the proposed SFI-LoRa scheme with perfect orthogonality assumption only have trivial gaps (less than $0.1$ dB) to their corresponding simulated SER results with imperfect orthogonality in the low-SNR region.
This not only validates the accuracy of our theoretical derivation in Section~IV, but also verifies the rationality of our assumption in the theoretical derivation.

\subsection{BER Performance}

In Fig.~5, the BER performance of the proposed SFI-LoRa scheme and the conventional LoRa, MuLoRa, DCDSK, HM-LoRa, and FBI-LoRa schemes over the AWGN, Rayleigh fading, and multipath flat-fading AWGN channels are presented. The discrete-time multipath flat-fading AWGN channel model is given by \cite{ref96,ref97}
\begin{equation}\label{flat-fading}
  r(n) = x_0(n) + \sum_{i=1}^{N_\mathrm{path}-1}h_\mathrm{m}x_i(n-n_i)+z(n),
\end{equation}
where $N_\mathrm{path}$, $n_i$ and $h_m$ are the number of paths, the discrete path delay, and the attenuation, respectively. In particular, the parameters of the six LoRa schemes are set as in Section~IV-B to
guarantee almost the same average symbol durations.
As shown in Fig.~5(a), compared with the conventional LoRa scheme, there is around $0.3~\rm{dB}$ performance loss for the proposed SFI-LoRa scheme over an AWGN channel. However, the BER performance of the SFI-LoRa scheme has performance gains of 0.8, 0.35, 1.3, $1.1~\rm{dB}$ compared to MuLoRa, DCDSK, HM-LoRa, and FBI-LoRa schemes at a BER of $10^{-4}$, respectively. Similar observations can also be obtained in Fig.~5(b) and Fig.~5(c) over the Rayleigh fading and multipath flat-fading AWGN channels. For example, in these two scenario, there are around 0.3 and $0.8~\rm{dB}$ performance loss for the proposed SFI-LoRa scheme compared with the conventional LoRa scheme, respectively. Moreover, the BER performance of the proposed SFI-LoRa scheme is higher than those of the MuLoRa (0.7 and 1.8~$\rm{dB}$), DCDSK (0.4 and 0.6~$\rm{dB}$), HM-LoRa (0.9 and 1.25~$\rm{dB}$), and FBI-LoRa schemes (1.3 and 2.1~$\rm{dB}$), respectively over the Rayleigh fading and multipath flat-fading AWGN channels.



\subsection{Throughput Performance}

Fig.~\ref{fig:throughput-compare} illustrates the throughput performance of the proposed SFI-LoRa, conventional LoRa, MuLoRa, DCDSK, HM-LoRa and FBI-LoRa schemes over AWGN, Rayleigh fading, and multipath flat-fading AWGN channels. 
As shown in Fig.~\ref{fig:throughput-compare}(a), the throughput performance of the proposed SFI-LoRa scheme is 170\%, 52\% , 35\%and 14\% higher than the conventional LoRa, MuLoRa, DCDSK, and HM-LoRa schemes in high-SNR region, respectively. Moreover, the throughput performance of the FBI-LoRa scheme is 14.8\% higher than the SFI-LoRa scheme. Similar observations can be also obtained in Figs.~\ref{fig:throughput-compare}(b) and \ref{fig:throughput-compare}(c) over the Rayleigh fading and multipath flat-fading AWGN channels, respectively. Referring to the Figs.~\ref{fig:throughput-compare}(b) and ~\ref{fig:throughput-compare}(c), although the throughput performance of the FBI-LoRa scheme in high-SNR region is 16\% and 5\% higher than the SFI-LoRa scheme, respectively.
The throughput performance of the proposed SFI-LoRa scheme is higher than those of the conventional LoRa (211\% and 168\%), MuLoRa (75\% and 51\%), DCDSK (56\% and 34\%), and HM-LoRa (12\% and 7\%) schemes, respectively.
To elaborate a little further, we provide a comparison of the throughput performance loss and BER performance gain between the SFI-LoRa scheme and the state-of-the-art LoRa schemes over a Rayleigh fading channel with a target BER of $10^{-4}$. With respect to the FBI-LoRa scheme, the performance gain of the proposed SFI-LoRa scheme is $\left(10^{\frac{S_{\mathrm{gap}}}{10}}-1\right) = 34.9\%$ \cite{ref11}, where $S_\mathrm{gap}=1.3~{\rm{dB}}$ is the SNR gap between the SFI-LoRa scheme and FBI-LoRa scheme achieving the target BER performance (see Fig.~\ref{fig:throughput-compare}(b)). In other words, the FBI-LoRa scheme requires about 34.9\% higher SNR than the proposed SFI-LoRa scheme to achieve the target BER performance. 
Obviously, compared with the FBI-LoRa scheme, the BER performance gain of the proposed SFI-LoRa scheme (34.9\%) is much greater than the corresponding throughput performance loss ($16\%$). On the other hand, compared to the conventional LoRa scheme, the proposed SFI-LoRa scheme obtains a desirable throughput performance gain (211\%) with a trivial BER performance loss (4\%). Moreover, compared with  MuLoRa, DCDSK, and HM-LoRa schemes, the proposed SFI-LoRa scheme not only has BER performance gains of 0.7, 0.4, and $0.9~{\rm{dB}}$, but also achieves throughput performance gains of  75\%, 56\%, and 12\%, respectively.

{\em Remark 3:} We have also carried out simulations and found that the available highest throughput of the proposed SFI-LoRa scheme (i.e., $M = 5$) is much larger than that of the conventional LoRa scheme (i.e., $\mathrm{SF}=7$).


%

\subsection{Effect of CFO and STO on BER Performance}

\begin{figure}[t]
\centering
\includegraphics[width=3.3 in]{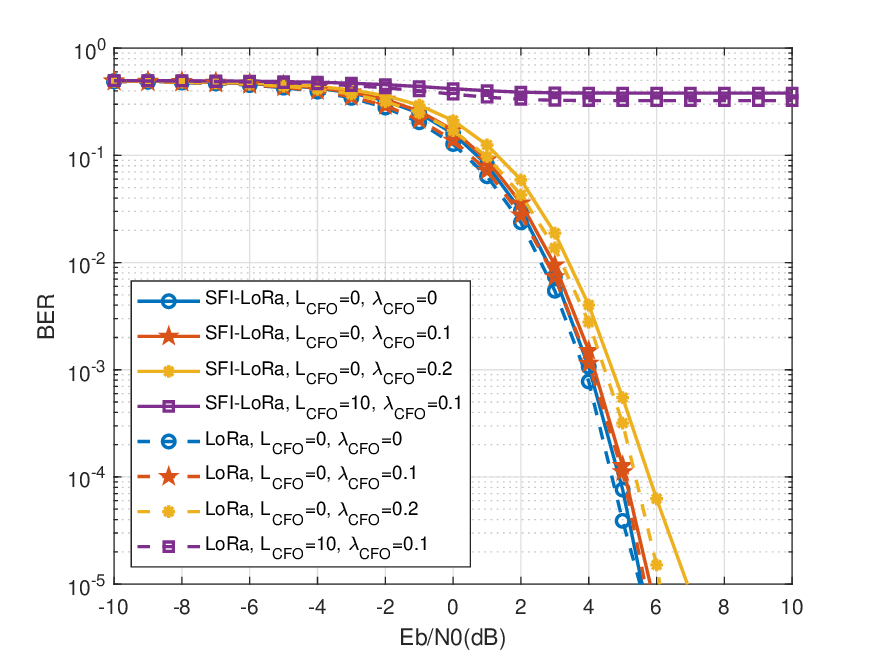}\vspace{-1mm}
\caption{Simulated BER results of the proposed SFI-LoRa over different CFO over an AWGN channel.}
\label{offset2}\vspace{-2mm}
\end{figure}

\begin{figure}[t]
\centering
\includegraphics[width=3.3 in]{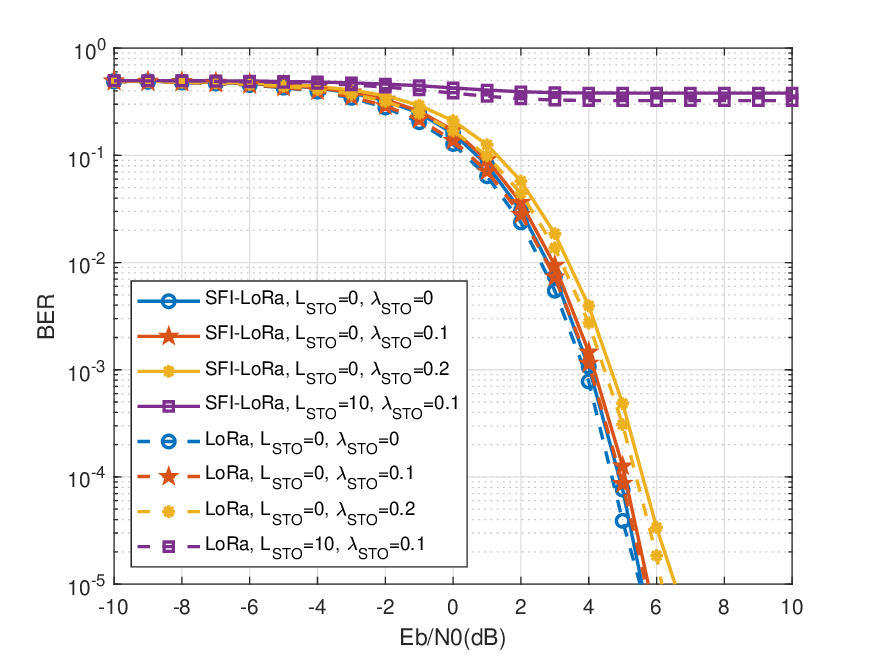}
\caption{Simulated BER results of the proposed SFI-LoRa over different STO over an AWGN channel.}
\label{offset3}
\end{figure}

In this subsection, we evaluate the effect of carrier frequency offset (CFO) and sampling time offset (STO) on the proposed SFI-LoRa scheme over an AWGN channel. According to the \cite{ref71}, the received signal with CFO and STO in the SFI-LoRa scheme can be denoted as
\begin{equation}y(t)=e^{j2\pi t\Delta f_c}x(t+\tau_t).\end{equation}
The CFO and STO can be respectively denoted as
\begin{equation}\Delta f_c=B_\mathrm{w}\cdot\frac{L_\mathrm{CFO}+\lambda_\mathrm{CFO}}{2^f},\end{equation}
\begin{equation}\tau_{\mathrm{t}}=\frac{L_{\mathrm{STO}}+\lambda_{\mathrm{STO}}}{B_{\mathrm{w}}},\end{equation}
 where $L_\mathrm{CFO}$ and $L_\mathrm{STO}$ are integers, $\lambda_\mathrm{CFO}$ and $\lambda_\mathrm{STO} \in [-0.5,0.5]$ are fractions. Since $L_\mathrm{CFO}$ and $L_\mathrm{STO}$ are integers, they shift an integer number of DFT bins. Moreover, the fractional offsets $\lambda_\mathrm{CFO}$ and $\lambda_\mathrm{STO}$ spread the energy of the symbol previously contained in a frequency bin over several frequency bins. Figs.~\ref{offset2} and \ref{offset3} illustrate the BER performance of the SFI-LoRa scheme with different offsets.
 When $L_\mathrm{CFO}$ or $L_\mathrm{STO}$ is not zero, it means that a serious offset occurs \cite{ref84}. In this scenario, the SFI-LoRa scheme cannot work without offset compensation (see Figs.~\ref{offset2} and \ref{offset3}). The similar phenomenon can also be observed in the conventional LoRa scheme.
 Moreover, as shown in Fig.~\ref{offset2}, the BER performance gap between the proposed scheme and the conventional LoRa scheme gradually widens as the CFO increases. When the CFO equals 0, 0.1, and 0.2, the performance gaps between the proposed scheme and the conventional LoRa scheme are 0.2 dB, 0.35 dB, and 0.6 dB, respectively. A similar phenomenon can also be observed in Fig.~\ref{offset3}. We believe that this phenomenon occurs because the transmission energy of the proposed SFI-LoRa scheme is spread to multiple LoRa signals with different SFs. Although our design is more sensitive to CFO and STO than the conventional LoRa scheme, this issue also exists in other variants of LoRa schemes \cite{ref79,ref80}. Nevertheless, in application scenarios requiring higher throughput \cite{ref96,ref81}, the proposed SFI-LoRa scheme can better satisfy the demand for high data rate.

\begin{table}[t]\label{tab5}
\renewcommand\arraystretch{1.2}\vspace{-1mm}
\caption{\textsc{Four Modes of Hata Path Loss Model.}}
\begin{tabular}{|c|c|c|}

\hline
Type of Area & \(a(h_\mathrm{r})\)                                                                                                                                          & \(K\)     \\ \hline
Open         & \multirow{3}{*}{\begin{tabular}[c]{@{}c@{}}\([1.1\mathrm{Log}_{10}(f_{\mathrm{MHz}})-0.7]h_{m}\)\\ \(-[1.56\mathrm{Log}_{10}(f_{\mathrm{Mhz}})-0.8]\)\end{tabular}}
& \begin{tabular}[c]{@{}c@{}}\(4.78[\mathrm{Log_{10}}(f_{\mathrm{Mhz}})]^{2}\)\\\(-18.33\mathrm{Log_{10}}(f_{\mathrm{Mhz}})\)
\\\(+40.94\)\end{tabular} \\ \cline{1-1} \cline{3-3}
Suburban     &                                                                                                                                                & \begin{tabular}[c]{@{}c@{}}\(2[\mathrm{Log}_{10}(f_{\mathrm{MHz}}/28)]^{2}\)\\\(+5.4\)\end{tabular} \\ \cline{1-1} \cline{3-3}
Small City   &                                                                                                                                                & 0     \\ \hline
Large City   & \begin{tabular}[c]{@{}c@{}}\(3.2[\mathrm{Log}_{10}(11.75h_\mathrm{r})]^{2}\)\\\(-4.97\) \end{tabular}                                                                                                                                         & 0     \\ \hline
\end{tabular}
\end{table}

\subsection{Coverage Probability}

In this subsection, the Hata path loss model \cite{ref16} is assumed to analyze the coverage probability of the proposed SFI-LoRa scheme. The Hata path loss model considers antenna height, propagation distance, frequency, and environment type, and thus has been widely used in the performance evaluation of LoRa schemes \cite{ref74,ref108,ref109}. In the Hata model, path loss is the most variable quantity in the link budget. It depends on frequency, antenna height, receive terminal location relative to obstacles and reflectors, and link distance, among many other factors \cite{ref16}. Specifically, the Hata path loss model has four modes: open, suburban, small city, and large city, as shown in Table~\ref{tab5}. Furthermore, the basic function for Hata path loss is derived as
\begin{equation}\begin{aligned}
L_{\mathrm{hata}}=&69.55+26.16\mathrm{Log}_{10}(f_{\mathrm{MHz}})-13.82\mathrm{Log}_{10}(h_{b})
\\&\!-\!a(h_{m})\!+\![44.9-6.55\mathrm{Log}_{10}(h_{b})]\mathrm{Log}_{10}(d_{\mathrm{km}})\!-\!K,
\end{aligned}\end{equation}
where \(h_\mathrm{b}\) and \(h_\mathrm{r}\) are the transmitting antenna height and receiving antenna height, respectively. Assuming that the transmitting antenna height \(h_\mathrm{b}=30\)m, receiving antenna height \(h_\mathrm{r}=1.5\)m and transmit power $P_t=14$ dBm.

Furthermore, the coverage probability of the proposed SFI-LoRa scheme is defined as $P_\mathrm{cov}(d)=\mathrm{Pr}(P_\mathrm{r}(d)\geq P_\mathrm{th})$, where $P_\mathrm{th}$ is the receiver sensitivity, $P_\mathrm{r}(d)$ is the received power and represented by \cite{ref108}
\begin{equation}\label{prd}
  P_\mathrm{r}(d)=P_\mathrm{t}+G_\mathrm{t}+G_\mathrm{r}-L_\mathrm{hata}(d)-\psi.
\end{equation}
In (40), $\psi$ follows a normal random distribution with mean $0$ and variance $\sigma^2$, $P_\mathrm{t}, G_\mathrm{t}$, and $G_\mathrm{r}$ are the transmitted power, transmit antenna gain, and receive antenna gain, respectively, which are set the same as in \cite{ref109}. Therefore, the received signal of the proposed SFI-LoRa scheme in the Hata path loss model over a Rayleigh fading channel can be expressed by \cite{ref108}
\begin{equation}
  r(n) = \sqrt{h10^{-\frac{P_\mathrm{r}(d)}{10}}}x(n)+z(n),
\end{equation}
where $\sqrt{h}$ is the Rayleigh fading gain, and the channel reduces to an AWGN channel if $h=1$.
Based on the discussion, one can easily obtain the coverage probability of the proposed SFI-LoRa scheme in the Hata path loss model via Monte Carlo simulations.

\begin{figure}[t]
\centering
\includegraphics[width=3.3 in]{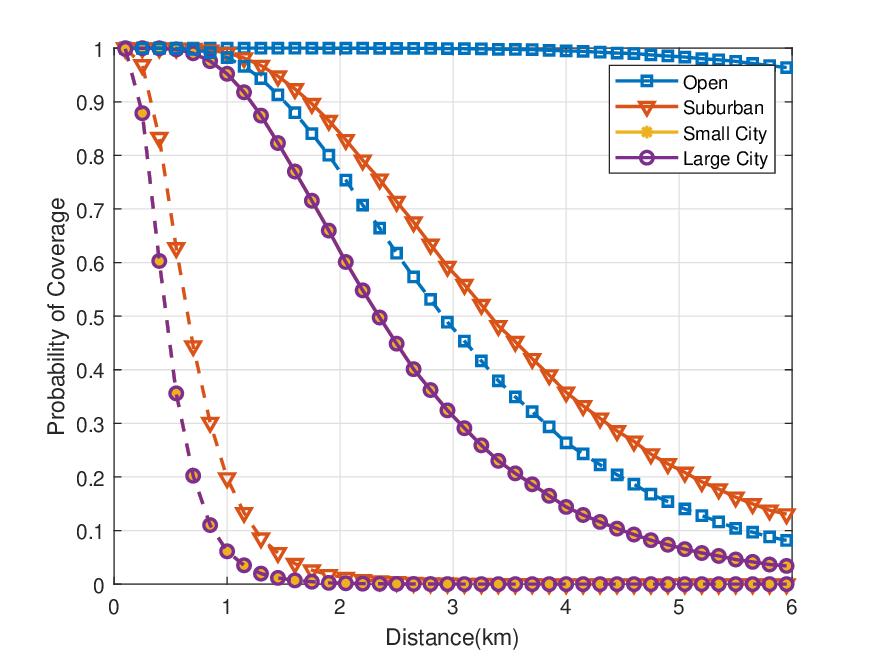}
\caption{ Coverage probability of the proposed SFI-LoRa scheme (\(M=2\)) for four modes (i.e., open, suburban, small city, and large city) of Hata path loss model over AWGN (solid line) and Rayleigh fading (dotted line) channels.}
\label{fig7}
\end{figure}
\begin{figure}[t]
\centering
\includegraphics[width=3.3 in]{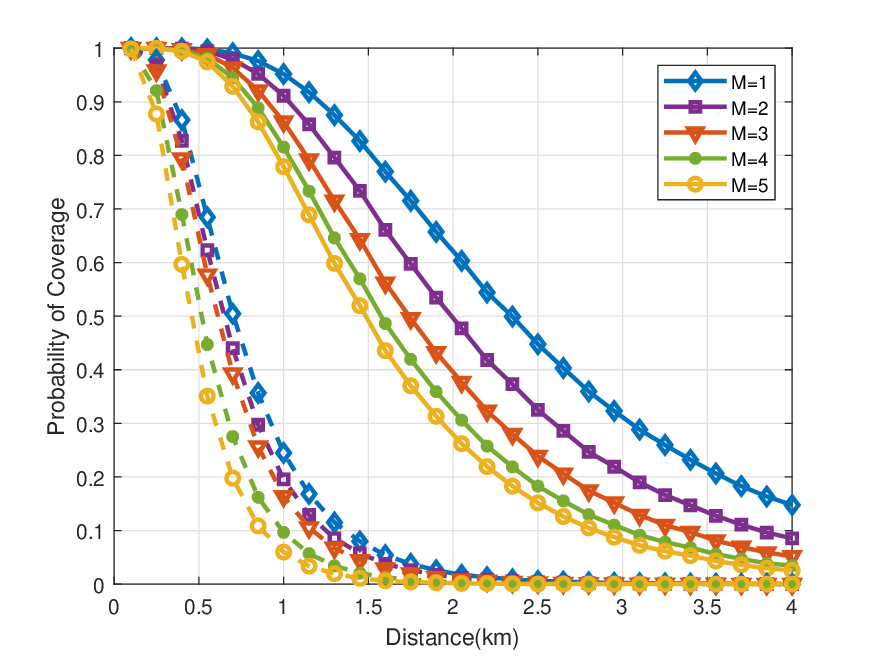}
\caption{Coverage probability of the proposed SFI-LoRa scheme for large city mode of Hata path loss model over AWGN (solid line) and Rayleigh fading (dotted line) channels.}
\label{fig8}
\end{figure}

According to \cite{ref16}, we define the SNR corresponding to the BER of $10^{-5}$ as the SNR threshold for successful decoding (i.e., receiver sensitivity). In this paper, we use this SNR threshold to evaluate the coverage probability of the SFI-LoRa scheme via Monte Carlo simulations.
Fig.~\ref{fig7} shows the coverage probability of the proposed SFI-LoRa scheme at different modes  based on the Hata path loss model over AWGN and Rayleigh fading channels.
We can observe that the transmission scenarios have a significant impact on coverage probability of the proposed SFI-LoRa scheme over AWGN and Rayleigh fading channels. Similar observations can be also obtained in existing LoRa systems \cite{ref69,ref70}. It can be observed that the relationship among the coverage areas of the four modes of the SFI-LoRa scheme is: large city $<$ small city $<$ suburbs $<$ open.

Fig.~\ref{fig8} compares the coverage of the proposed SFI-LoRa scheme under different numbers of selected SFs. It can be seen that the coverage gradually decreases as $M$ increases. Nevertheless, the proposed SFI-LoRa system can still maintain desirable coverage performance by adjusting $M$. For example, over an AWGN channel, when $M = 3$ and the coverage probability is 0.9, the data rate is 6.518 kbps and the distance is around 1.05 km. When $M = 5$ under the same coverage probability, the data rate increases to 12.146 kbps and the distance decreases to 0.82 km. In other words, as $M$ varies from $3$ to $5$, the data rate increases by 86.3\% (from 6.518 kbps to 12.146 kbps), while the transmission distance only decreases by 21.9\% (from 1.05 km to 0.82 km). Therefore, the proposed SFI-LoRa scheme can improve the throughput performance with a slight reduction of transmission distance. Similar observation can be also found over a Rayleigh fading channel. 
Moreover, according to Fig.~\ref{fig7} and Fig.~\ref{fig8}, the coverage probability of the proposed scheme is related to the scenario (i.e., open, suberban, small city, or large city), channels (i.e. AWGN channel or Rayleigh channel), and the throughput performance (i.e. the value of $M$). In practical applications, one should properly select these key parameters to achieve a desirable coverage probability.

\section{Conclusion}

This paper explores the design and performance analysis of a novel SFI-LoRa scheme for realizing high-data rate transmissions. To be specific, we use the indices of SF combinations as a new signal dimension to achieve this goal. Moreover, the proposed SFI-LoRa scheme can transmit additional information bits through a combinatorial method by incorporating an index mapping module at the transmitter. The information bits are modulated by both the combination of the SF indices and SFB. In addition, the SER analyses and simulations have been conducted to validate the accuracy of our theoretical derivation. Furthermore, we have carefully compared the proposed SFI-LoRa with conventional LoRa, MuLoRa, DCDSK, HM-LoRa, and FBI-LoRa schemes in terms of BER performance, data rate, transmission throughput, complexity and energy efficiency to demonstrate the superiority of our design. As a consequence, the proposed SFI-LoRa scheme is regarded as a potentially valuable candidate for practical IoT applications characterized with high data-rate requirements. In the future, we will explore the joint design of channel coding and the SFI-LoRa scheme as well as evaluate its practical performance using experimental platform.


\end{document}